\newif\ifsubmission
\submissionfalse
\documentclass[11pt,letterpaper]{article}
\usepackage[margin=1in]{geometry}

\usepackage{amsmath,amsthm,amssymb}
\usepackage{xspace}
\usepackage{url}
\usepackage{hyperref}
\usepackage[dvipsnames]{xcolor}
\usepackage{algorithm}
\usepackage{algpseudocode}
\usepackage{appendix}

\usepackage{multirow}
\usepackage{tablefootnote}
\usepackage{booktabs}
\usepackage[capitalise,noabbrev]{cleveref}

\hypersetup{colorlinks=true,allcolors=blue}

\theoremstyle{plain}
\newtheorem{theorem}{Theorem}[section]
\newtheorem{lemma}[theorem]{Lemma}
\newtheorem{claim}[theorem]{Claim}
\newtheorem{corollary}[theorem]{Corollary}
\newtheorem{question}[theorem]{Question}

\newtheorem{fact}[theorem]{Fact}

\theoremstyle{definition}
\newtheorem{definition}[theorem]{Definition}

\theoremstyle{remark}
\newtheorem{remark}[theorem]{Remark}

\newcommand{\ProblemName}[1]{\textsc{#1}}
\newcommand{\kzC}{\ProblemName{$(k, z)$-Clustering}\xspace}
\newcommand{\kMedian}{\ProblemName{$k$-Median}\xspace}
\newcommand{\kMeans}{\ProblemName{$k$-Means}\xspace}

\newcommand{\JL}{Johnson-Lindenstrauss\xspace}

\newcommand{\calF}{\ensuremath{\mathcal{F}}\xspace}
\newcommand{\calG}{\ensuremath{\mathcal{G}}\xspace}
\newcommand{\calH}{\ensuremath{\mathcal{H}}\xspace}

\newcommand{\calP}{\ensuremath{\mathcal{P}}\xspace}

\newcommand{\hatV}{\ensuremath{\hat V}\xspace}
\newcommand{\hatP}{\ensuremath{\hat P}\xspace}

\newcommand{\RR}{{\mathbb R}}
\DeclareMathOperator{\E}{\mathbb{E}}

\DeclareMathOperator{\cost}{cost}
\newcommand{\sdim}{\ensuremath{\mathrm{sdim}}\xspace}
\newcommand{\hdim}{\ensuremath{\mathrm{hdim}}\xspace}

\DeclareMathOperator{\poly}{poly}
\newcommand{\tw}{\ensuremath{\mathrm{tw}}\xspace}
\newcommand{\OPT}{\ensuremath{\mathrm{OPT}}\xspace}

\DeclareMathOperator{\NN}{NN}
\newcommand{\minn}[1]{\min\{{#1}\}}

\providecommand{\set}[1]{{\{#1\}}}
\newcommand{\Capx}{{C^\mathrm{apx}}}
\newcommand{\Xapx}{X^\mathrm{apx}}
\newcommand{\sigmaapx}{\sigma^\mathrm{apx}}
\newcommand{\Cmax}{{C^\mathrm{max}}}

\def\compactify{\itemsep=0pt \topsep=0pt \partopsep=0pt \parsep=0pt}

\newcommand*\samethanks[1][\value{footnote}]{\footnotemark[#1]}

\title{Coresets for Clustering in Excluded-minor Graphs and Beyond}
\author{Vladimir Braverman\thanks{Johns Hopkins University.
Email: \texttt{\{vova@cs.jhu.edu, xwu71@jh.edu\}}
} \and
Shaofeng H.-C. Jiang\thanks{Weizmann Institute of Science.
  Work partially supported by ONR Award N00014-18-1-2364,
  the Israel Science Foundation grant \#1086/18,
  and a Minerva Foundation grant.
  Part of this work was done while some of the authors were visiting the Simons Institute for the Theory of Computing.
Email: \texttt{\{shaofeng.jiang, robert.krauthgamer\}@weizmann.ac.il}} \and
Robert Krauthgamer\samethanks \and
Xuan Wu\samethanks[1]
}
\date{}

\begin{document}

\setcounter{page}{0}

\begin{titlepage}
    \maketitle
    \thispagestyle{empty}
    \begin{abstract}
Coresets are modern data-reduction tools
that are widely used in data analysis to improve efficiency
in terms of running time, space and communication complexity.
Our main result is a fast algorithm to construct a small coreset
for \kMedian in (the shortest-path metric of) an excluded-minor graph.
Specifically, we give the first coreset of size that depends only on $k$, $\epsilon$ and the excluded-minor size,
and our running time is quasi-linear (in the size of the input graph).

The main innovation in our new algorithm is that is iterative;
it first reduces the $n$ input points to roughly $O(\log n)$ reweighted points,
then to $O(\log\log n)$, and so forth until the size is independent of $n$.
Each step in this iterative size reduction is based on the
importance sampling framework of Feldman and Langberg (STOC 2011),
with a crucial adaptation that reduces the number of \emph{distinct points},
by employing a terminal embedding
(where low distortion is guaranteed only for the distance
from every terminal to all other points).
Our terminal embedding is technically involved
and relies on shortest-path separators,
a standard tool in planar and excluded-minor graphs.

Furthermore, our new algorithm is applicable also in Euclidean metrics,
by simply using a recent terminal embedding result of Narayanan and Nelson, (STOC 2019), which extends the Johnson-Lindenstrauss Lemma.
We thus obtain an efficient coreset construction in high-dimensional Euclidean spaces, thereby matching and simplifying state-of-the-art results
(Sohler and Woodruff, FOCS 2018; Huang and Vishnoi, STOC 2020).

In addition, we also employ terminal embedding with additive distortion
to obtain small coresets in graphs with bounded highway dimension,
and use applications of our coresets to obtain improved approximation schemes, e.g., an improved PTAS for planar \kMedian via a new centroid set. \end{abstract}

 \end{titlepage}

\newpage
\ifsubmission
\else

\section{Introduction}
\label{sec:intro}

Coresets are modern tools for efficient data analysis that have become widely used in theoretical computer science, machine learning, networking and other areas.
This paper investigates coresets for the metric \kMedian problem that can be defined as follows.
Given an \emph{ambient} metric space $M=(V, d)$
and a \emph{weighted} set $X \subseteq V$ with weight function $w : X \to \mathbb{R}_+$,
the goal is to find a set of $k$ \emph{centers} $C \subseteq V$ that minimizes the total cost of connecting every point to a center in $C$:
\begin{align*}
    \cost(X, C) := \sum_{x \in X}{w(x) \cdot d(x, C)},
\end{align*}
where $d(x, C) := \min_{y \in C}{d(x, y)}$ is the distance to the closest center.
An \emph{$\epsilon$-coreset} for \kMedian on $X$ is a weighted subset $D\subseteq X$, such that
\begin{align*}
    \forall C \subseteq V, |C| = k , \qquad \cost(D, C) \in (1\pm \epsilon) \cdot \cost(X, C).
\end{align*}
We note that many papers study a more general problem, \kzC, where inside the cost function each distance is raised to power $z$.
We focus on \kMedian for sake of exposition, but most of our results easily extend to \kzC.

Small coresets are attractive since one can solve the problem on $D$ instead of $X$ and, as a result, improve time, space or communication complexity of downstream applications \cite{liang2013distributed, lucic2017training,fss13}.
Thus, one of the most important performance measures of a coreset $D$ is its \emph{size}, i.e., the number of distinct points in it, denoted $\| D \|_0$.\footnote{For a weighted set $X$,
  we denote by $\|X\|_0$ the number of \emph{distinct} elements,
  by $\|X\|_1$ its total weight.
}
Har-Peled and Mazumdar~\cite{HM04} introduced the above definition
and designed the first coreset for \kMedian in Euclidean spaces
($V=\RR^m$ with $\ell_2$ norm),
and since their work,
designing small coresets has become a flourishing research direction,
including not only \kMedian and \kzC e.g.~\cite{DBLP:journals/dcg/Har-PeledK07,Chen09,DBLP:conf/soda/LangbergS10,DBLP:conf/stoc/FeldmanL11,DBLP:conf/focs/SohlerW18,HV20,fss13},
but also many other important problems, such as
subspace approximation/PCA~\cite{FFM06,FMSW10,fss13},
projective clustering~\cite{DBLP:conf/stoc/FeldmanL11, DBLP:conf/fsttcs/VaradarajanX12, fss13},
regression~\cite{maalouf2019fast},
density estimation~\cite{DBLP:conf/colt/KarninL19,phillips2019near},
ordered weighted clustering~\cite{DBLP:conf/icml/BravermanJKW19},
and fair clustering~\cite{DBLP:conf/waoa/0001SS19,DBLP:conf/nips/HuangJV19}.

Many modern coreset constructions stem from a fundamental framework
proposed by Feldman and Langberg~\cite{DBLP:conf/stoc/FeldmanL11},
extending the importance sampling approach of Langberg and Schulman~\cite{DBLP:conf/soda/LangbergS10}.
In this framework~\cite{DBLP:conf/stoc/FeldmanL11},
the size of an $\epsilon$-coreset for \kMedian is bounded by $O(\poly(k/\epsilon)\cdot \sdim)$,
where $\sdim$ is the shattering (or VC) dimension of the family of distance functions.
For a general metric space $(V,d)$,
a direct application of \cite{DBLP:conf/stoc/FeldmanL11} results
in a coreset of size $O_{k,\epsilon}(\log |V|)$,
which is tight in the sense that in some instances,
every coreset must have size $\Omega(\log |V|)$
\cite{coreset_tw}.Therefore, to obtain coresets of size independent of the data set $X$,
we have to restrict our attention to specific metric spaces,
which raises the following fundamental question.

\begin{question}\label{question1}
Identify conditions on a data set $X$ from metric space $(V,d)$
that guarantee the existence (and efficient construction) of
an $\epsilon$-coreset for \kMedian of size $O_{\epsilon,k}(1)$?
\end{question}

This question has seen major advances recently.
Coresets of size independent of $X$ (and $V$) were obtained,
including efficient algorithms, for several important special cases:
high-dimensional Euclidean spaces
\cite{DBLP:conf/focs/SohlerW18, DBLP:journals/corr/abs-1912-12003, HV20}
(i.e., independently of the Euclidean dimension),
metrics with bounded doubling dimension
\cite{DBLP:conf/focs/HuangJLW18},
and shortest-path metric of bounded-treewidth graphs
\cite{coreset_tw}.

\subsection{Our Results}

\paragraph{Overview}
We make significant progress on this front (Question~\ref{question1})
by designing new coresets for \kMedian
in three very different types of metric spaces.
Specifically, we give
(i) the first $O_{\epsilon, k}(1)$-size coreset for excluded-minor graphs;
(ii) the first $O_{\epsilon, k}(1)$-size coreset for graphs with bounded highway dimension; and
(iii) a simplified state-of-the-art coreset for high-dimensional Euclidean spaces (i.e., coreset-size independent of the Euclidean dimension
with guarantees comparable to \cite{HV20} but simpler analysis.)

Our coreset constructions are all based on the well-known importance sampling
framework of~\cite{DBLP:conf/stoc/FeldmanL11},
but with subtle deviations that introduce significant advantages.
Our first technical idea is to relax the goal of computing the final coreset
in one shot:
we present a general reduction that turns an algorithm
that computes a coreset of size $O(\poly(k/\epsilon) \log{\|X\|_0})$
into an algorithm that computes a coreset of size $O(\poly(k/\epsilon))$.
The reduction is very simple and efficient, by straightforward iterations.
Thus, it suffices to construct a coreset of size
$O(\poly(k / \epsilon) \log{ \|X\|_0 } )$.
We construct this using the importance sampling framework~\cite{DBLP:conf/stoc/FeldmanL11},
but applied in a subtly different way, called terminal embedding,
in which distances are slightly distorted,
trading accuracy for (hopefully) a small shattering dimension.
It still remains to bound the shattering dimension,
but we are now much better equipped ---
we can distort the distances (design a new embedding or employ a known one),
and we are content with dimension bound $O_{k,\epsilon}(\log\|X\|_0)$,
instead of the usual $O_{k,\epsilon}(1)$.

We proceed to present each of our results and its context-specific background,
see also \cref{tab:result} for summary,
and then describe our techniques at a high-level
in \cref{sec:tech_contrib}.

\begin{table}[ht]
    \centering
    \caption{our results of $\epsilon$-coresets for \kMedian in various
    types of metric spaces $M(V, d)$ with comparison to previous works.
    By graph metric, we mean the shortest-path metric of an edge-weighted graph $G = (V, E)$.
    \cref{cor:coreset_euclidean} (and~\cite{HV20}) also work for general \kzC, but we list the result for \kMedian ($z = 1$) only.
    }
    \begin{tabular}[t]{clll}
        \toprule
        \multicolumn{2}{c}{
        Metric space} & Coreset size\tablefootnote{Throughout, the notation $\tilde O(f)$ hides $\poly\log f$ factors,
  and $O_m(f)$ hides factors that depend on $m$. 
} & Reference \\
        \midrule
\multicolumn{2}{c}{
        General metrics} & $\tilde{O}(\epsilon^{-2}k\log |V|)$ & \cite{DBLP:conf/stoc/FeldmanL11} \\
\cmidrule{1-2}
        \multirow{3}{*}{Graph metrics} & Bounded treewidth & $\tilde{O}(\epsilon^{-2}k^3 )$ & \cite{coreset_tw} \\
& Excluding a fixed minor
        & $\tilde{O}(\epsilon^{-4}k^2)$ & \cref{cor:coreset_mf}\\
& Bounded highway dimension
        & $\tilde{O}(k^{O(\log(1/\epsilon))})$ & \cref{cor:coreset_hw} \\
        \cmidrule{1-2}
        \multirow{2}{*}{Euclidean $\mathbb{R}^{m}$} & Dimension-dependent
        & $\tilde{O}(\epsilon^{-2} k m)$ & \cite{DBLP:conf/stoc/FeldmanL11} \\
        & Dimension-free & 
        $\tilde{O}(\epsilon^{-4}k)$
        & \cite{HV20}, \cref{cor:coreset_euclidean} \\
\bottomrule
\end{tabular}
    \label{tab:result}
\end{table}

\paragraph{Coresets for Clustering in Graph Metrics}
\kMedian clustering in graph metrics, i.e. shortest-path metric of graphs, is a central task in data mining of spatial networks (e.g., planar networks such as road networks)~\cite{DBLP:journals/tkde/ShekharL97,DBLP:conf/sigmod/YiuM04},
and has applications in various location optimization problems,
such as placing servers on the Internet~\cite{DBLP:conf/infocom/LiGIDS99,DBLP:conf/infocom/JaminJJRSZ00} (see also a survey~\cite{tansel1983state}),
and in data analysis methods~\cite{DBLP:conf/icml/RattiganMJ07,DBLP:journals/tvcg/CuiZQWL08}.
We obtain new coresets for excluded-minor graphs and new coresets for graphs of bounded highway dimension.
The former generalize planar graphs and the latter capture the structure of transportation networks.

\paragraph{Coresets for Excluded-minor Graphs}
A \emph{minor} of graph $G$ is a graph $H$ obtained from $G$
by a sequence of edge deletions, vertex deletions or edge contractions.
We are interested in graphs $G$ that exclude a fixed graph $H$ as a minor,
i.e., they do not contain $H$ as a minor.
Excluded-minor graphs have found numerous applications in theoretical computer science and beyond and they include, for example, planar graphs and bounded-treewidth graphs.
Besides its practical importance, \kMedian in planar graphs
received significant attention in approximation algorithms research~\cite{Thorup05,cohen2019local,DBLP:conf/esa/Cohen-AddadPP19}.
Our framework yields the first $\epsilon$-coreset of size $O_{k,\epsilon}(1)$ for \kMedian in excluded-minor graphs,
see \cref{cor:coreset_mf} for details.
Such a bound was previously known only for the special case
of bounded-treewidth graphs~\cite{coreset_tw}.
We stress that our technical approach is significantly different from~\cite{coreset_tw}; 
we introduce a novel iterative construction
and a relaxed terminal embedding of excluded-minor graph metrics
(see \cref{sec:tech_contrib}), 
and overall bypass bounding the shattering dimension by $O(1)$
(which is the technical core in~\cite{coreset_tw}).

\paragraph{Coresets for Graphs with Bounded Highway Dimension}
Due to the tight relation to road networks, graphs of bounded highway dimension is another important family for the study of clustering in graph metrics.
The notion of highway dimension was first proposed by~\cite{DBLP:conf/soda/AbrahamFGW10}
to measure the complexity of transportation networks such as road networks and airline networks.
Intuitively, it captures the fact that going from any two far-away
cities $A$ and $B$, the shortest path between $A$ and $B$ always goes through
a small number of connecting hub cities. The formal definition of highway dimension is given in \cref{def:hdim}, and we compare
related versions of definitions in \cref{remark:highway}.
The study of highway dimension was originally to understand the
efficiency of heuristics for shortest path computations~\cite{DBLP:conf/soda/AbrahamFGW10},
while subsequent works also study approximation algorithms for optimization problems such as TSP, Steiner Tree~\cite{FFKP18} and \kMedian~\cite{DBLP:conf/esa/BeckerKS18}.
We show the first coreset for graphs with bounded highway dimension,
and as we will discuss later it can be applied to design new approximation algorithms.
The formal statement can be found in \cref{cor:coreset_hw}.

\paragraph{Coresets for High-dimensional Euclidean Space}
The study of coresets for \kMedian (and more generally \kzC) in Euclidean space $\mathbb{R}^{m}$
spans a rich line of research. The first coreset for \kMedian
in Euclidean spaces, given by~\cite{HM04}, has size $O(k \epsilon^{-m} \log n)$ where $n=\|X\|_1$,
and the $\log n$ factor was shaved by a subsequent work~\cite{DBLP:journals/dcg/Har-PeledK07}.
The exponential dependence on the Euclidean dimension $m$ was later improved
to $\poly(km / \epsilon)$ \cite{DBLP:conf/soda/LangbergS10},
and to $O(k m / \epsilon^{2} )$ \cite{DBLP:conf/stoc/FeldmanL11}.
Very recently, the first coreset for \kMedian of size $\poly(k/\epsilon)$,
which is \emph{independent} of the Euclidean dimension $m$,\footnote{Dimension-independent coresets were obtained earlier
  for Euclidean \kMeans~\cite{DBLP:journals/corr/BravermanFL16,fss13},
  however these do not apply to \kMedian.
}
was obtained by~\cite{DBLP:conf/focs/SohlerW18}
(see also~\cite{DBLP:journals/corr/abs-1912-12003}).\footnote{The focus of~\cite{DBLP:conf/focs/SohlerW18} is on \kMedian,
  but the results extend to \kzC.
}
This was recently improved in~\cite{HV20},
which designs a (much faster) near-linear time construction for \kzC,
with slight improvements in the coreset size
and the (often useful) additional property that the coreset is a subset of $X$.
Our result extends this line of research;
an easy application of our new framework yields
a near-linear time construction of coreset of size $\poly(k/\epsilon)$,
which too is independent of the dimension $m$.
Compared to the state of the art~\cite{HV20}, our result achieves essentially the same size bound, while greatly simplifying the analysis. A formal statement and detailed comparison with~\cite{HV20} can be found in \cref{cor:coreset_euclidean} and \cref{remark:euclidean_coreset}.

\paragraph{{Applications: Improved Approximation Schemes}}
We apply our coresets to design approximation schemes for \kMedian in
shortest-path metrics of planar graphs and graphs with bounded highway dimension.
In particular, we give an FPT-PTAS, parameterized by $k$ and $\epsilon$,
in graphs with bounded highway dimension (\cref{cor:fpt_ptas_hw}),
and a PTAS in planar graphs (\cref{cor:planar_ptas}).
Both algorithms run in time near-linear in $|V|$, and improve previous
results in the corresponding settings.

The PTAS for \kMedian in planar graphs is obtained using a new centroid-set result.
A \emph{centroid set} is a subset of $V$ that contains centers giving a
$(1 + \epsilon)$-approximate solution.
We obtain centroid sets of size \emph{independent} of the input $X$ in planar graphs,
which improves a recent size bound $(\log{|V|})^{O(1/\epsilon)}$  \cite{DBLP:conf/esa/Cohen-AddadPP19},
and moreover runs in time near-linear in $|V|$.
This centroid set can be found in \cref{thm:centroid_planar}.

\subsection{Technical Contributions}
\label{sec:tech_contrib}

\paragraph{Iterative Size Reduction}
This technique is based on an idea so simple that it may seem too naive:
Basic coreset constructions have size $O_{k,\epsilon}(\log n)$,
so why not apply it repeatedly, to obtain a coreset
of size $O_{k, \epsilon}(\log \log{n})$, then $O_{k, \epsilon}(\log \log \log n)$ and so on?
One specific example is the size bound $O(\epsilon^{-2} k \log n)$
for a general $n$-point metric space \cite{DBLP:conf/stoc/FeldmanL11},
where this does not work because
$n=|V|$ is actually the size of the \emph{ambient} space,
irrespective of the \emph{data} set $X$.
Another example is the size bound $O(\epsilon^{-m} k \log n)$
for Euclidean space $\RR^m$ \cite{HM04},
where this does not work because
$n=\|X\|_1$ is the total weight of the data points $X$,
which coresets do not reduce (to the contrast, they maintain it).
These examples suggest that one should avoid two pitfalls:
dependence on $V$ and dependence on the total weight.

We indeed make this approach work by requiring an algorithm $\mathcal{A}$
that constructs a coreset of size $O(\log{\|X\|_0})$, which is \emph{data-dependent} (recall that $\|X\|_0$ is the number of \emph{distinct}
elements in a weighted set $X$).
Specifically, we show in \cref{thm:ite_size_reduct} that,
given an algorithm $\mathcal{A}$ that
constructs an $\epsilon'$-coreset of size $O(\poly(k/\epsilon')\log{\|X\|_0})$
for every $\epsilon'$ and $X \subseteq V$,
one can obtain an $\epsilon$-coreset of size $\poly(k/\epsilon)$
by simply applying $\mathcal{A}$ iteratively.
It follows by setting $\epsilon'$ carefully,
so that it increases quickly and eventually $\epsilon'=O(\epsilon)$.
See \cref{sec:SizeReduction} for details.

Not surprisingly, the general idea of applying the sketching/coreset algorithm iteratively
was also used in other related contexts (e.g.~\cite{DBLP:conf/focs/LiMP13,DBLP:conf/soda/ClarksonW15,DBLP:conf/nips/MunteanuSSW18}).
Moreover, a related two-step iterative construction
was applied in a recent coreset result~\cite{HV20}.
Nevertheless,
the exact implementation of iterative size reduction in coresets is unique in the literature.
As can be seen from our results, this reduction fundamentally
helps to achieve new or simplified coresets of size \emph{independent} of data set.
We expect the iterative size reduction to be of independent interest to future research.

\paragraph{Terminal Embeddings}
To employ the iterative size reduction, we need to
construct coresets of size $\poly(k/\epsilon)\cdot \log{\| X \|_0}$.
Unfortunately, a direct application of~\cite{DBLP:conf/stoc/FeldmanL11} yields a bound that depends on the number of vertices $|V|$, irrespective of $X$.
To bypass this limitation, the framework of~\cite{DBLP:conf/stoc/FeldmanL11} is augmented (in fact, we use a refined framework proposed in~\cite{fss13}),
to support controlled modifications to the distances $d(\cdot, \cdot)$.
As explained more formally in \cref{sec:generalized_fl},
one represents these modifications using
a set of functions $\calF = \{ f_x : V \to \mathbb{R}_+ \mid x \in X \}$,
that corresponds to the modified distances from each $x$,
i.e., $f_x(\cdot) \leftrightarrow d(x,\cdot)$.
Many previous papers~\cite{DBLP:conf/soda/LangbergS10, DBLP:conf/stoc/FeldmanL11, DBLP:journals/corr/BravermanFL16, fss13}
work directly with the distances and use the function set
$\calF = \{ f_x(\cdot) = d(x, \cdot) \mid x \in X \}$,
or a more sophisticated but still direct variant of hyperbolic balls
(where each $f_x$ is an affine transformation of $d(x,\cdot)$).
A key difference is that we use a ``proxy'' function set $\calF$,
where each $f_x(\cdot) \approx d(x,\cdot)$.
This introduces a tradeoff between the approximation error (called distortion)
and the shattering dimension of $\calF$ (which controls the number of samples),
and overall results in a smaller coreset.
Such tradeoff was first used in~\cite{DBLP:conf/focs/HuangJLW18}
to obtain small coresets for doubling spaces,
and was recently used in~\cite{HV20} to reduce the coreset size for Euclidean spaces.
This proxy function set may be alternatively viewed as
a \emph{terminal embedding} on $X$,
in which both the distortion of distances (between $X$ and all of $V$)
and the shattering dimension are controlled.

We then consider two types of terminal embeddings $\calF$.
The first type (\cref{sec:TEmultiplicative})
maintains $(1+\epsilon)$-multiplicative distortion of the distances.
When this embedding achieves dimension bound $O(\poly(k/\epsilon) \log{\|X\|_0})$,
we combine it with the aforementioned iterative size reduction,
to further reduce the size to be independent of $X$.
It remains to actually design embeddings of this type,
which we achieve (as explained further below),
for excluded-minor graphs and for Euclidean spaces,
and thus we overall obtain $O_{\epsilon, k}(1)$-size coresets in both settings.
Our second type of terminal embeddings $\calF$ (\cref{sec:TEadditive})
maintains additive distortion on top of the multiplicative one.
We design embeddings of this type (as explained further below)
for graphs with bounded highway dimension;
these embeddings have shattering dimension $\poly(k/\epsilon)$,
and thus we overall obtain $O_{\epsilon, k}(1)$-size coresets
even without the iterative size reduction.
We report our new terminal embeddings in \cref{tab:new_sdim}.

\begin{table}[ht]
    \centering
    \caption{New terminal embeddings $\calF$ for different metrics spaces.
    The reported distortion bound is the upper bound on $f_x(c)$,
    in addition to the lower bound $f_x(c) \geq d(x, c)$.
    The embeddings of graphs with bounded highway dimension,
    called here ``highway graphs'' for short,
    are defined with respect to a given $S \subseteq V$ (see \cref{lemma:hw_sdim}).
    }
    \begin{tabular}[t]{llll}
        \toprule
        Metric space & Dimension $\sdim_{\max}(\calF)$ & Distortion & Result \\
        \midrule
        Euclidean & $O(\epsilon^{-2}\log\|X\|_0)$ & $(1 + \epsilon)\cdot d(x, c)$ & \cref{lemma:euclidean_em} \\
Excluded-minor graphs & $\tilde{O}(\epsilon^{-2}\log\|X\|_0)$ & $(1 + \epsilon)\cdot d(x, c)$ & \cref{lemma:mf_sdim} \\
Highway graphs & $O(|S|^{O(\log(1/\epsilon))})$ &
        $(1 + \epsilon) \cdot d(x, c) + \epsilon \cdot d(x, S)$ & \cref{lemma:hw_sdim} \\
        \bottomrule
    \end{tabular}
    \label{tab:new_sdim}
\end{table}

\paragraph{Terminal Embedding for Euclidean Spaces}
Our terminal embedding for Euclidean spaces is surprisingly simple,
and is a great showcase for our new framework.
In a classical result~\cite{DBLP:conf/stoc/FeldmanL11},
it has been shown that $\sdim_{\max}(\calF) = O(m)$
for Euclidean distance in $\mathbb{R}^{m}$ without distortion.
On the other hand, we notice a terminal embedding version
of Johnson-Lindenstrauss Lemma
was discovered recently~\cite{DBLP:conf/stoc/NarayananN19}.
Our terminal embedding bound (\cref{lemma:euclidean_em})
follows by directly combining these two results,
see \cref{sec:Euclidean} for details.

We note that without our iterative size reduction technique,
plugging in the recent terminal Johnson-Lindenstrauss  Lemma~\cite{DBLP:conf/stoc/NarayananN19}
into classical importance sampling frameworks,
such as~\cite{DBLP:conf/stoc/FeldmanL11,fss13}
does not yield any interesting coreset.
Furthermore, the new terminal Johnson-Lindenstrauss Lemma was recently used
in~\cite{HV20} to design coresets for high-dimensional Euclidean spaces.
Their size bounds are essentially the same as ours,
however they go through a complicated analysis to directly show
a shattering dimension bound $\poly(k / \epsilon)$.
This complication is not necessary in our method,
because by our iterative size reduction it suffices to show
a very loose $O_{k,\epsilon}(\log \|X\|_0)$ dimension bound,
and this follows immediately from the Johnson-Lindenstrauss result.

\paragraph{Terminal Embedding for Excluded-minor Graphs}
The technical core of the terminal embedding for excluded-minor
graphs is how to bound the shattering dimension.
In our proof, we reduce the problem of bounding the shattering dimension
into finding a representation of the distance functions
on $X \times V$ as a set of \emph{min-linear} functions.
Specifically, we need to find for each $x$
a min-linear function $g_x : \mathbb{R}^s \to \mathbb{R}$ of the form $g_x(t) = \min_{1 \leq i \leq s}\{a_i t_i + b_i\}$,
where $s = O(\log{\|X\|_0})$, such that $\forall c \in V$, there
is $t \in \mathbb{R}^s$ with $d(x, c) = g_x(t)$.

The central challenge is how to relate the graph structure to
the structure of shortest paths $d(x, c)$.
To demonstrate how we relate them,
we start with discussing the simple special case of bounded treewidth graphs.
For bounded treewidth graphs, the vertex separator theorem is applied to find
a subset $P \subseteq V$, through which the shortest path $x \rightsquigarrow y$ has to pass.
This translates into the following
\begin{align*}
    d(x, c) = \min_{p \in P}\{ d(x, p) + d(p, c) \},
\end{align*}
and for each $x \in X$, we can use this to define
the desired min-linear function
$g_x(d(p_1, c), \ldots, d(p_m, c))$ $= d(x, c)$, where we write $P = \{p_1, \ldots, p_m\}$.

However, excluded-minor graphs do not have small vertex separator,
and we use the shortest-path separator~\cite{DBLP:journals/jacm/Thorup04,DBLP:conf/podc/AbrahamG06} instead.  
Now assume for simplicity that the shortest paths $x \rightsquigarrow c$
all pass through a fixed shortest path $l$.
Because $l$ itself is a shortest path, we know
\begin{align*}
    \forall x \in X, c\in V, \quad
    d(x, c) = \min_{u_1, u_2 \in l}\{d(x, u_1) + d(u_1, u_2) + d(u_2, c)\}.
\end{align*}
Since $l$ can have many (i.e. $\omega(\log{\|X\|_0})$) points,
we need to discretize $l$ by designating $\poly(\epsilon^{-1})$ \emph{portals}
$P^l_x$ on $l$ for each $x \in X$ (and similarly $P^l_c$ for $c \in V$).
This only introduces $(1 + \epsilon)$ distortion to the distance, which we can afford.

Then we create $d'_x : l \to \mathbb{R}_+$ to approximate
$d(x,u)$'s, using distances from $x$ to the portals $P^l_x$ (and similarly for $d(c, u)$).
Specifically, for the sake of presentation, assume $P^l_x = \{p_1, p_2, p_3\}$ ($p_1 \leq p_2 \leq p_3$), interpret $l$ as interval $[0, 1)$,
then for $u \in [0, p_1)$, define $d'_x(u) =  d(x, 0)$,
for $u \in [p_1, p_2)$, define $d'_x(u) = d(x, p_1)$, and so forth.
Hence, each $d'_x(\cdot)$ is a piece-wise linear function of $O(|P^l_x|)$ pieces (again, similarly for $d'_c(\cdot)$), and this enables us to write
\begin{align*}
    d(x, c) \approx d'(x, c) := \min_{u_1, u_2 \in P^l_x \cup P^l_c}\{ d'_x(u_1) + d(u_1, u_2) + d'_c(u_2) \}.
\end{align*}

Therefore, it suffices to find a min-linear representation for $d'(x, \cdot)$ for $x \in X$.
However, the piece-wise linear structure of $d'_x$ creates extra difficulty to define min-linear representations.
To see this, still assume $P^l_x = \{p_1, p_2, p_3\}$.
Then to determine $d'_x(u)$ for $ u \in P^l_x \cup P^l_c$,
we not only need to know $d(x, p_i)$ for $p_i \in P^l_x$,
but also need to know
which sub-interval $[p_i, p_{i+1})$ that $u$ belongs to.
(That is, if $u \in [p_1, p_2)$, then $d'_x(u) = d(x, p_1)$.)
Hence, in addition to using distances $\{c\} \times P^l_c$ as variables of $g_x$,
the relative ordering between points in $P^l_x \cup P^l_c$
is also necessary to evaluate $d'(x, c)$.

Because $c \in V$ can be arbitrary, we cannot simply ``remember''
the ordering in $g_x$. Hence, we ``guess'' this ordering, and for each fixed ordering we can write $g_x$ as a min-linear function of few variables.
Luckily, we can afford the ``guess'' since $|P^l_x \cup P^l_c| = \poly(\epsilon^{-1})$ which is independent of $X$.
A more detailed overview can be found in \cref{sec:ExcludedMinor}.

\paragraph{Terminal Embedding for Graphs with Bounded Highway Dimension}
In addition to a $(1 + \epsilon)$ multiplicative error,
the embedding for graphs with bounded highway dimension
also introduces an additive error.
In particular, for a given $S \subseteq V$,
it guarantees that
\begin{align*}
    \forall x \in X, c \in V,\quad d(x, c) \leq f_x(c) \leq
    (1 + \epsilon) \cdot d(x, c) + \epsilon \cdot d(x, S).
\end{align*}
This terminal embedding is a direct consequence of a similar
embedding from graphs with bounded highway dimension to graphs with
bounded treewidth~\cite{DBLP:conf/esa/BeckerKS18},
and a previous result about the shattering dimension for
graphs with bounded treewidth~\cite{coreset_tw}.
In our applications, we will choose $S$ to be a constant approximate
solution\footnote{in fact, a bi-criteria approximation suffices.} $C^\star$ to \kMedian.
So the additive error becomes $\epsilon \cdot d(x, C^\star)$.
In general, this term can still be much larger than $d(x, c)$,
but the \emph{collectively} error in the clustering objective is bounded.
This observation helps us to obtain a coreset, and due to the additional
additive error, the shattering dimension is already independent
of $X$ and hence no iterative size reduction is necessary.

 \subsection{Related Work}

Approximation algorithms for metric \kMedian have been extensively studied.
In general metric spaces, it is NP-hard to approximate \kMedian
within a $1+\frac{2}{e}$ factor~\cite{jain2002new},
and the state of the art is a $(2.675+\epsilon)$-approximation~\cite{byrka2014improved}.
In Euclidean space $\mathbb{R}^{m}$,
\kMedian is APX-hard if both $k$ and the dimension $m$ are part of the input~\cite{guruswami2003embeddings}.
However, PTAS's do exist if either $k$ or dimension $m$ is fixed~\cite{HM04,arora1998approximation,cohen2019local,DBLP:journals/siamcomp/FriggstadRS19}.

Tightly related to coresets, dimensionality reduction has also been studied for clustering in Euclidean spaces.
Compared with coresets which reduce the data set size while keeping the dimension,
dimensionality reduction
aims to find a low-dimensional representation of data points (but not necessarily reduce the number of data points).
As a staring point, a trivial application of Johnson-Lindenstrauss Lemma~\cite{JL84} yields a dimension bound $O(\epsilon^{-2} \log n)$ for \kzC.
For \kMeans with $1+\epsilon$ approximation ratio, \cite{cohen2015dimensionality} showed an $O(k/\epsilon^2)$ dimension bound
for data-oblivious dimension reduction and an $O(k/\epsilon)$ bound for the data-dependent setting.
Moreover, the same work~\cite{cohen2015dimensionality} also obtained
a data-oblivious $O(\epsilon^{-2}\log k)$ dimension bound for \kMeans with
approximation ratio $9+\epsilon$.
Very recently, \cite{becchetti2019oblivious} obtained an
$\tilde{O}(\epsilon^{-6}(\log k+\log \log n))$ dimension bound for \kMeans
and \cite{makarychev2019performance} obtained an
$O(\epsilon^{-2}\log\frac{k}{\epsilon})$ bound for \kzC.
Both of them used data-oblivious methods and have approximation ratio $1+\epsilon$.
Dimensionality reduction techniques are also used for constructing dimension-free coresets in Euclidean spaces \cite{DBLP:conf/focs/SohlerW18,becchetti2019oblivious,HV20,fss13}.

 \section{Preliminaries}
\label{sec:prelim}

\paragraph{Notations}
Let $V^k := \{ C \subseteq V : |C| \leq k \}$ denote the collection of all subsets of $V$ of size at most $k$.
\footnote{Strictly speaking, $V^k$ is the collection of all ordered $k$-tuples of $V$, but here we use it to denote the subsets.
Note that tuples may contain repeated elements so the subsets in $V^k$ are of size at most $k$.}
For integer $n, i > 0$, let $\log^{(i)}n$ denote the $i$-th iterated logarithm of $n$, i.e. $\log^{(1)} n := \log n$
and $\log^{(i)} n := \log(\log^{(i - 1)}{n})$ ($i \geq 2$).
Define $\log^\star n$ as the number of times the logarithm is iteratively applied before the result is at most $1$, i.e. $\log^\star n := 0$ if $n \leq 1$ and $\log^\star n = 1 + \log^\star(\log n)$ if $n > 1$.
For a weighted set $S$, denote the weight function as $w_S : S \to \mathbb{R}_+$.
Let $\OPT_z(X)$ be the optimal objective value for \kzC on $X$,
and we call a subset $C \subseteq V$ an $(\alpha,\beta)$-approximate solution for \kzC on $X$ if $|C| = \alpha k$
and $\cost_z(X, C) := \sum_{x \in X}{w_X(x) \cdot (d(x, C))^z} \leq \beta \cdot \OPT_z(X)$.

\paragraph{Functional Representation of Distances}
We consider sets of functions $\calF$ from $V$ to $\mathbb{R}_+$.
Specifically, we consider function sets $\calF = \{ f_x : V \to \mathbb{R}_+ \mid x \in X\}$ that is indexed by the weighted data set $X \subseteq V$,
and intuitively
$f_x(\cdot)$
is used to measure the distance from $x \in X$ to a point in $V$.
Because we interpret $f_x$'s as distances, for a subset $C \subseteq V$, we define $f_x(C) := \min_{c \in C}{f_x(C)}$,
and define the clustering objective accordingly as
\begin{align*}
    \cost_z(\calF, C) := \sum_{f_x \in \calF}{w_\calF(f_x) \cdot (f_x(C)})^z.
\end{align*}
In fact, in our applications, we will use $f_x(y)$ as a ``close'' approximation to $d$.
We note that this functional representation is natural for $k$-Clustering,
since the objective function only uses distances from $X$ to every $k$-subset of $V$ only.
Furthermore, we do not require the triangle inequality to hold for such functional representations.

\paragraph{Shattering Dimension}
For $c \in V, r\geq 0$, define $B_\calF(c, r) := \{ f \in \calF : f(c) \leq r \}$.
We emphasize that $c$ is from the ambient space $V$ in addition to the data set $X$.
Intuitively, $B_\calF(c, r)$ is the ball centered at $c$ with radius $r$
when the $f$ functions are used to measure distances. 
For example, consider $X = V$ and let $f_x(\cdot) := d(x, \cdot)$ for $x\in V$.
Then $B_\calF(c, r) = \{f_x \in \calF  : d(c, x) \leq r \}$, which corresponds to the metric ball centered at $c$ with radius $r$.

We introduce the notion of shattering dimension in \cref{def:sdim}.
In fact, the shattering dimension may be defined with respect to
any set system~\cite{har2011geometric},
but we do not need this generality here and thus
we consider only the shattering dimension of the ``metric balls'' system.

\begin{definition}[Shattering Dimension~\cite{har2011geometric}]
    \label{def:sdim}
    Suppose $\calF$ is a set of functions from $V$ to $\mathbb{R}_+$.
    The shattering dimension of $\calF$, denoted as $\sdim(\calF)$, is the smallest integer $t$, such that for every $\calH \subseteq \calF$ with $|\calH| \geq 2$, 
    \begin{equation}
        \forall \calH \subseteq \calF, |\calH| \geq 2, \quad \left| \{ B_\calH(c, r) : c \in V, r \geq 0 \} \right| \leq |\calH|^t. \label{eqn:sdim}
    \end{equation}
\end{definition}

The shattering dimension is tightly related to the well-known VC-dimension~\cite{vapnik1971uniform},
and they are equal to each other up to a logarithmic factor~\cite[Corollary 5.12, Lemma 5.14]{har2011geometric}.
In our application, we usually do not use $\sdim(\calF)$ directly.
Instead, given a point weight $v : X \to \mathbb{R}_+$,
we define $\calF_v := \{ f_x \cdot v(x) \mid x \in X \}$,
and then consider the maximum of $\sdim(\calF_v)$ over all possible $v$,
defined as $\sdim_{\max}(\calF) := \max_{v : X \to \mathbb{R}_+}{ \sdim(\calF_v) }$.

 \section{Framework}
\label{sec:framework}
We present our general framework for constructing coresets.
Our first new idea is a generic reduction, called iterative size reduction,
through which it suffices to find a coreset of size $O(\log {\|X\|_0})$
only in order to get a coreset of size independent of $X$.
This general reduction greatly simplifies the coreset construction,
and in particular, as we will see, ``old'' techniques such as importance sampling gains new power and becomes useful for new settings such as excluded-minor graphs.

Roughly speaking, the iterative size reduction
turns a coreset construction algorithm $\mathcal{A}(X, \epsilon)$ with size $O(\poly(\epsilon^{-1}k)\cdot\log\|X\|_0)$
into a construction $\mathcal{A}'(X, \epsilon)$ with size $\poly(\epsilon^{-1} k)$.
To define $\mathcal{A}'$, we simply iteratively apply $\mathcal{A}$,
i.e. $X_i := \mathcal{A}(X_{i-1}, \epsilon_i)$, and terminate when $\|X_i\|_0$ does not decrease.
However, if $\mathcal{A}$ is applied for $t$ times in total,
the error of the resulted coreset is accumulated as $\sum_{i = 1}^{t}{\epsilon_t}$.
Hence, to make the error bounded, we make sure $\epsilon_i \geq 2\epsilon_{i-1}$
and $\epsilon_t = O(\epsilon)$,
so $\sum_{i=1}^{t}{\epsilon_i} = O(\epsilon)$.
Moreover, our choice of $\epsilon_i$ also guarantees that $\|X_i\|_0$
is roughly $\poly(\epsilon^{-1}k \cdot \log^{(i)}{\|X\|_0})$.
Since $\log^{(i)}{\|X\|_0}$ decreases very fast with respect to $i$,
$\|X_i\|_0$ becomes $\poly(\epsilon^{-1} k)$ in about $t = \log^\star{\|X\|_0}$ iterations.
The detailed algorithm $\mathcal{A}'$ can be found in \cref{alg:ite_size_reduct}, and we present the formal analysis in \cref{thm:ite_size_reduct}.

To construct the actual coresets which is to be used with the reduction,
we adapt the importance sampling method that was proposed by Feldman and Langberg~\cite{DBLP:conf/stoc/FeldmanL11}.
In previous works, the size of the coresets from importance sampling
is related to the shattering dimension of metric balls system (i.e. in our language, it is the shattering dimension of $\calF = \{ d(x, \cdot) \mid x \in X \}$.)
Instead of considering the metric balls only,
we give a generalized analysis where we consider a general set of ``distance functions'' $\calF$
that has some error but is still ``close'' to $d$.
The advantage of doing so is that we could trade the accuracy with
the shattering dimension, which in turn reduces the size of the coreset.

We particularly examine two types of such functions $\calF = \{ f_x : V \to \mathbb{R}_+ \mid x \in X \}$.
The first type $\calF$ introduces a multiplicative $(1 + \epsilon)$ error
to $d$, i.e. $\forall x \in X, c \in V$, $d(x, c) \leq f_x(c) \leq (1 + \epsilon) \cdot d(x, c)$.
Such a
small distortion is already very helpful to obtain an $O(\log{\|X\|_0})$ shattering dimension
for minor-free graphs and Euclidean spaces.
In addition to the multiplicative error, the other type of $\calF$
introduces a certain additive error, and we make use of this
to show $O(k)$ shattering dimension bound for bounded highway dimension graphs and doubling spaces.
In this section, we will discuss how the two types of function sets
imply efficient coresets, and the dimension bounds for various metric families will be analyzed in \cref{sec:coresets} where we also present the coreset results.

\subsection{Iterative Size Reduction}
\label{sec:SizeReduction}

\begin{theorem}[Iterative Size Reduction]
    \label{thm:ite_size_reduct}
    Let $\rho \geq 1$ be a constant and let $\mathcal{M}$ be a family of metric spaces.
    Assume $\mathcal{A}(X, k, z, \epsilon, \delta, M)$ is a randomized algorithm that
    constructs an $\epsilon$-coreset of size
    $\epsilon^{-\rho} s(k) \log{\delta^{-1}} \log{\|X \|_0}$ for \kzC
    on every weighted set $X \subseteq V$ and metric space $M(V, d) \in \mathcal{M}$,
    for every $z \geq 1, 0 < \epsilon,\delta < \frac{1}{4}$,
    running in time $T(\|X\|_0, k, z, \epsilon, \delta, M)$
    with success probability $1 - \delta$.
    Then algorithm $\mathcal{A}'(X, k, z, \epsilon, \delta, M)$,
    stated in \cref{alg:ite_size_reduct},
    computes an $\epsilon$-coreset of size $\tilde{O}(\epsilon^{-\rho} s(k)  \log {\delta^{-1}})$
    for \kzC on every weighted set $X \subseteq V$ and metric space $M(V, d) \in \mathcal{M}$, for every $z \geq 1$, $0 < \epsilon, \delta < \frac{1}{4}$,
    in time
    \begin{align*}
        O(T(\|X\|_0, k, z, O(\epsilon/(\log \|X\|_0)^{\frac{1}{\rho}}), O(\delta/\|X\|_0), M) \cdot \log^\star{\|X\|_0}),
    \end{align*}
    and with success probability $1 - \delta$.
\end{theorem}

\begin{algorithm}[ht]
    \caption{Iterative size reduction $\mathcal{A}'(X, k, z, \epsilon, \delta, M)$}
    \label{alg:ite_size_reduct}
    \begin{algorithmic}[1]
        \Require algorithm $\mathcal{A}(X, k, z, \epsilon, \delta, M)$
        that computes an $\epsilon$-coreset for \kzC on $X$
        with size $\epsilon^{-\rho} s(k) \log{\delta^{-1}} \log{\| X \|_0}$
        and success probability $1 - \delta$.
        \State let $X_0 := X$, and let $t$ be the largest integer such that
        $\log^{(t-1)}{\|X\|_0} \geq \max\{20\epsilon^{-\rho} s(k)\log{\delta^{-1}},\rho 2^{\rho+1}\}$
        \For{$i = 1,\cdots,t$}
            \State let $\epsilon_i := \epsilon / (\log^{(i)}{\| X \|_0})^{\frac{1}{\rho}}$,
            $\delta_i := \delta/\|X_{i-1}\|_0$
            \State let $X_{i} := \mathcal{A}(X_{i-1}, k, z, \epsilon_i, \delta_i, M)$
        \EndFor
        \State $X_{t+1}:= \mathcal{A}(X_t, k, z, \epsilon, \delta, M)$
        \State return $X_{t+1}$
    \end{algorithmic}
\end{algorithm}
\begin{proof}
    For the sake of presentation, let $n := \| X \|_0$, $s := s(k)$,
    and $\Gamma := s \epsilon^{-\rho} \log{\delta^{-1}}$.
    We start with proving in the following that $X_t$ is an $O(\epsilon)$-coreset of $X$ with size $\max\{160000\Gamma^4,20\Gamma\rho^3 2^{3\rho+3}\}$ with probability $1-O(\delta)$.

    Let $a_i := \|X_i\|_0$. Then by definition of $X_i$,
    \begin{align}
      a_i&=s\epsilon_i^{-\rho}\log a_{i-1}\log\delta_i^{-1} \nonumber\\
      &=s\epsilon_i^{-\rho} \log a_{i-1} (\log a_{i-1}+\log \delta^{-1})\nonumber \\
      &\leq s\epsilon_i^{-\rho}\log \delta^{-1}(\log a_{i-1})^2 \label{eqn:a_i_ub}
\end{align}
    where the inequality is by $\log a_{i-1}+\log \delta^{-1}\leq \log a_{i-1}\cdot \log \delta^{-1}$,
    which is equivalent to $(\log a_{i-1}-1)(\log\delta^{-1}-1)\geq 1$
    and the latter is true because $a_{i-1}\geq \epsilon^{-\rho} \geq \epsilon^{-1}\geq 4$ and $\delta<\frac{1}{4}$.

    Next we use induction to prove that $a_i\leq 20 \Gamma \log{\delta^{-1}} (\log^{(i)}n)^3$ for all $i=1,\ldots, t$.
    This is true for the base case when $i=1$, since $a_1\leq s\epsilon_1^{-\rho} \log{\delta^{-1}} (\log n)^2
    \leq \Gamma(\log n)^3 \leq 20\Gamma(\log n)^3$.
    Then consider the inductive case $i\geq 2$ and assume the hypothesis is true for $i-1$. We have
    \begin{align*}
    a_i
    &\leq s\epsilon_i^{-\rho}\log{\delta^{-1}}(\log a_{i-1})^2
      & \text{by \eqref{eqn:a_i_ub}} \\
    &= \Gamma \log^{(i)}{n}\cdot (\log a_{i-1})^2 & \text{by definition of $\epsilon_i$} \\
    &\leq \Gamma \log^{(i)}{n} \cdot (\log (20\Gamma (\log^{(i-1)}n)^3))^2
      & \text{by induction hypothesis}\\
    &= \Gamma \log^{(i)}{n} \cdot (\log (20\Gamma)+3\log^{(i)}n)^2\\
    &\leq \Gamma \log^{(i)}{n} \cdot (2(\log(20\Gamma))^2+18(\log^{(i)}n)^2)
      & \text{by $(a+b)^2\leq 2a^2+2b^2$}\\
    &\leq 20 \Gamma (\log^{(i)}n)^3,
    \end{align*}
    where the last inequality follows from the fact that
    $\log(20\Gamma )\leq \log(\log^{(i-1)} n)=\log^{(i)} n$, by $i \leq t$ and the definition of $t$.
Hence we conclude $a_i\leq 20\Gamma (\log^{(i)}n)^3$.
    This in particular implies that
    $a_t\leq 20\Gamma (\log^{(t)}n)^3$, and by definition of $t$, we have $\log^{(t)}n<\max\{20 \Gamma ,\rho 2^{\rho+1}\}$. Hence,
    \begin{align*}
        a_t\leq \max\{160000\Gamma^4,20\Gamma\rho^3 2^{3\rho+3}\}.
    \end{align*}
    By the guarantee of $\mathcal{A}$, we know that $X_t$ is a $\Pi_{i=1}^t (1+\epsilon_i)$-coreset for $X$. Note that $a\geq 2^\rho\log a$ for every $a\geq \rho 2^{\rho+1}$, so we have $\epsilon_{i+1}\geq 2\epsilon_i$ for $i\leq t$, which implies that $\sum_{i=1}^t \epsilon_i\leq 2\epsilon_t$. Hence we conclude that
    \begin{align*}
    \Pi_{i=1}^t (1+\epsilon_i)
    \leq \exp\left(\sum_{i=1}^t \epsilon_i\right)
    \leq \exp(2\epsilon_t)
    \leq \exp\left( \frac{2\epsilon}{(\log^{(t)}n)^{\frac{1}{\rho}}}\right)
    \leq \exp(2\epsilon)
    \leq 1 + 10 \epsilon,
    \end{align*}
    where the second last inequality follows from
    $\log^{(t)} n=\log (\log ^{(t-1)} n)\geq \log (\rho 2^{\rho+1})\geq 1$ for $\rho\geq 1$,
    and the last inequality follows by the fact that
    $\exp(2\epsilon)\leq 1+10\epsilon$ for $\epsilon\in (0,1)$.
    For the failure probability, we observe that $a_{i-1}\geq \epsilon_{i-1}^{-\rho}\geq \log^{(i-1)} n$, hence $\delta_i=\frac{\delta}{a_{i-1}}\leq \frac{\delta}{\log^{(i-1)} n}$, and the total failure probability is 
    \begin{align*}
    \sum_{i=1}^t \delta_i
    \leq \delta\left(\frac{1}{n}+\frac{1}{\log n}+\cdots+\frac{1}{\log ^{(t-1)} n}\right)
    \leq O(\delta),
    \end{align*}
    where again we have used $\log ^{(t-1)} n\geq \rho 2^{\rho+1}\geq 4$, by definition of $t$ and $\rho\geq 1$.
    
    Therefore, $X_t$ is an $O(\epsilon)$-coreset of $X$ with size
    $\max\{160000 \Gamma^4, 20\Gamma \rho^3 2^{3\rho+3}\}$ with probability $1-O(\delta)$.
    Finally, in the end of algorithm $\mathcal{A}'$,
    we apply $\mathcal{A}$ again on $X_t$ with parameter $\epsilon$ and $\delta$ to obtain an $O(\epsilon)$-coreset of $X$ with size $s\epsilon^{-\rho}\log\delta^{-1}\log (\max\{160000\Gamma^4,20\Gamma\rho^3 2^{3\rho+3}\})
=\tilde{O}(s\epsilon^{-\rho}\log{\delta^{-1}})$
    with probability $1-O(\delta)$.

    To see the running time, we note that $t=O(\log^\star n)$,
    and we run $\mathcal{A}$ for $t+1$ times. Moreover,
    since $\epsilon_i \geq \epsilon_1$ and $\delta_i \geq \delta_1$,
    the running time of each call of $\mathcal{A}$ is at most
    $T(\|X\|_0, k, z, \epsilon_1, \delta_1, M)$.
    This completes the proof of \cref{thm:ite_size_reduct}.
\end{proof}

\subsection{Importance Sampling}
\label{sec:generalized_fl}

We proceed to design the algorithm $\mathcal{A}$ required
by \cref{thm:ite_size_reduct}.
It is based on the importance sampling algorithm
introduced by~\cite{DBLP:conf/soda/LangbergS10,DBLP:conf/stoc/FeldmanL11},
and at a high level consists of two steps:
\begin{enumerate} \compactify
\item Computing probabilities: for each $x \in X$,
  compute $p_x\ge 0$ such that $\sum_{x \in X}{p_x} = 1$.
\item Sampling: draw $N$ (to be determined later) independent samples from $X$,
  each drawn from the distribution $(p_x : x \in X)$,
and assign each sample $x$ a weight $\frac{w_X(x)}{p_x\cdot N}$
  to form a coreset $D$.
\end{enumerate}

The key observation in the analysis of this algorithm
is that the sample size $N$, which is also the coreset size $\|D\|_0$,
is related to the shattering dimension (see \cref{def:sdim})
of a suitably defined set of functions~\cite[Theorem 4.1]{DBLP:conf/stoc/FeldmanL11}.
The analysis in~\cite{DBLP:conf/stoc/FeldmanL11} has been subsequently
improved~\cite{DBLP:journals/corr/BravermanFL16,fss13},
and we make use of~\cite[Theorem 31]{fss13},
restated as follows.

\begin{lemma}[Analysis of Importance Sampling~\cite{fss13}]
\label{lemma:generalized_fl}
Fix $z \geq 1$, $0 <  \epsilon  < \frac{1}{2}$,
an integer $k \geq 1$ and a metric space $(V, d)$.
Let $X \subseteq V$ have weights $w_X : V \to \mathbb{R}_+$
and let $\calF := \{ f_x : V \to \mathbb{R}_+ \mid x \in X \}$
be a corresponding set of functions with weights $w_{\calF}(f_x) = w_X(x)$.
Suppose $\{\sigma_x \}_{x \in X}$ satisfies
    \begin{align*}
      \forall x \in X,\quad
      \sigma_x
      \geq \sigma^\calF_x
      := \max_{C \in V^k}{ \frac{ w_X(x) \cdot (f_x(C))^z }{ \cost_z(\calF, C) } } ,
    \end{align*}
    and set a suitable
    \begin{align*}
        N = O(\epsilon^{-2}  \sigma_X  ( k \cdot \sdim_{\max}(\calF) \cdot \log(\sdim_{\max}(\calF)) \cdot \log{\sigma_X} + \log \tfrac{1}{\delta})),
    \end{align*}
    where $\sigma_X := \sum_{x \in X}{\sigma_x}$ and
    \begin{align*}
        \sdim_{\max}(\calF) := \max_{v : X \to \mathbb{R}_+}\sdim\left(\calF_v \right), \qquad \calF_v := \{ f_x \cdot v(x) \mid x \in X \}.
    \end{align*}
    Then the weighted set $D$ of size $\|D\|_0 = N$
    returned by the above importance sampling algorithm satisfies,
    with high probability $1 - \delta$,
    \begin{align*}
        \forall C \in V^k, \quad \sum_{x \in D}{w_D(x) \cdot (f_x(C))^z}
        \in (1\pm \epsilon) \cdot \cost_z(\calF, C).
    \end{align*}
\end{lemma}

\begin{remark}
We should explain how~\cite[Theorem 31]{fss13}
implies \cref{lemma:generalized_fl}.
First of all, the bound in~\cite{fss13} is with respect to VC-dimension,
and we transfer to shattering dimension by losing a logarithmic factor (see \cref{sec:prelim} for the relation between VC-dimension and shattering dimension).
Another main difference is that the functions therein are actually not from $V$ to $\mathbb{R}_+$.
For $\calF = \{ f_x : V \to \mathbb{R}_+ \mid x \in X \}$,
they consider $\calF^k := \{ f_x(C) = \min_{c \in C}\{f_x(c)\} \mid x \in X \}$,
and their bound on the sample size is
\begin{align*}
  N = \tilde{O}(\epsilon^{-2}  \sigma_X  ( \sdim_{\max}(\calF^k) \cdot \log{\sigma_X} + \log \tfrac{1}{\delta})).
\end{align*}
The notion of balls and shattering dimension they use (for $\calF^k$)
is the natural extension of our \cref{def:sdim}
(from functions on $V$ to functions on $V^k$),
where a ball around $C \in V^k$ is
$B_\calF(C, r) = \{ f_x \in \calF : f_x(C) \leq r \}$,
and~\eqref{eqn:sdim} is replaced by
\begin{align*}
  \left| \{ B_\calH(C, r) : C \in V^k, r \geq 0 \} \right| \leq |\calH|^t.
\end{align*}
Our \cref{lemma:generalized_fl}
follows from \cite[Theorem 31]{fss13}
by using the fact $\sdim(\calF^k) \leq k \cdot \sdim(\calF)$
from~\cite[Lemma 6.5]{DBLP:conf/stoc/FeldmanL11}.
\end{remark}

\paragraph{Terminal Embeddings.}
As mentioned in \cref{sec:intro}, $\calF$ in \cref{lemma:generalized_fl}
corresponds to the distance function $d$, i.e., $f_x(\cdot) = d(x,\cdot)$,
and \cref{lemma:generalized_fl} is usually applied directly to the distances,
i.e., on a function set $\calF = \{ f_x(\cdot) = d(x, \cdot) \mid x \in X \}$.
In our applications, we instead use \cref{lemma:generalized_fl}
with a ``proxy'' function set $\calF$ that is viewed
as a \emph{terminal embedding} on $X$,
in which both the distortion of distances (between $X$ and all of $V$)
and the shattering dimension are controlled.

We consider two types of terminal embeddings $\calF$.
The first type (\cref{sec:TEmultiplicative})
maintains $(1+\epsilon)$-multiplicative distortion of the distances,
and achieves dimension bound $O(\poly(k/\epsilon) \log{\|X\|_0})$,
and the other type of $\calF$ (\cref{sec:TEadditive})
maintains additive distortion on top of the multiplicative one,
but then the dimension is reduced to $\poly(k/\epsilon)$.
In what follows, we discuss how each type of terminal embedding
is used to construct coresets.

\subsection{Coresets via Terminal Embedding with Multiplicative Distortion}
\label{sec:TEmultiplicative}

The first type of terminal embedding distorts distances between $V$ and $X$
multiplicatively, i.e.,
\begin{align} \label{eq:MultDist}
  \forall x \in X, c \in V,
  \qquad
  d(x, c) \leq f_x(c) \leq (1 + \epsilon) \ d(x, c).
\end{align}
This natural guarantee works very well for \kzC in general.
In particular, using such $\calF$ in \cref{lemma:generalized_fl},
our importance sampling algorithm will produce (with high probability)
an $O(z\epsilon)$-coreset for \kzC.

\paragraph{Sensitivity Estimation.}
To compute a coreset using \cref{lemma:generalized_fl}
we need to define, for every $x\in X$,
\[
  \sigma_x \geq \sigma^\calF_x = \max_{C \in V^k}{ \frac{w_X(x)\cdot (f_x(C))^z }{ \cost_z(\calF, C) } } .
\]
The quantity $\sigma_x^\calF$,
usually called the \emph{sensitivity} of point $x\in X$ with respect to $\calF$
\cite{DBLP:conf/soda/LangbergS10,DBLP:conf/stoc/FeldmanL11};
essentially measures the maximal contribution of $x$
to the clustering objective over all possible centers $C \subseteq V$.
Since $f_x(y)$ approximates $d(x,y)$ by~\eqref{eq:MultDist},
it actually suffices to estimate the sensitivity with respect to $d$ instead of $\calF$, given by
\begin{align} \label{eq:sigmaStar}
  \sigma^\star_x
  := \max_{C \in V^k}{ \frac{w_X(x) \cdot (d(x, C))^z}{\cost_z(X, C)} }.
\end{align}

Even though computing $\sigma^\star_x$ exactly seems computationally difficult,
we shown next (in \cref{lemma:sensitivity}) that a good estimate
can be efficiently computed given an $(O(1), O(1))$-approximate clustering.
A weaker version of this lemma was presented in~\cite{DBLP:conf/fsttcs/VaradarajanX12}
for the case where $X$ has unit weights,
and we extend it to $X$ with general weights.
We will need the following notation.
Given a subset $C \subseteq V$,
denote the nearest neighbor of $x\in X$,
i.e., the point in $C$ closest to $x$ with ties broken arbitrarily,
by $\NN_C(x):= \arg\min \{d(x,y): y\in C \}$.
The tie-breaking guarantees that every $x$ has a unique nearest neighbor,
and thus $\NN_C(.)$ partitions $X$ into $|C|$ subsets.
The \emph{cluster of $x$ under $C$} is then defined as
$C(x) := \{ x' \in X : \NN_C(x') = \NN_C(x) \}$.

\begin{lemma}
\label{lemma:sensitivity}
Fix $z \geq 1$, an integer $k \geq 1$, and a weighted set $X$.
Given $\Capx \in V^k$ that is an $(\alpha,\beta)$-approximate solution
for \kzC on $X$,
define for every $x \in X$,
\begin{align*}
  \sigmaapx_x := w_X(x) \cdot \left(\frac{(d(x, \Capx))^z}{\cost_z(X, \Capx)} + \frac{1}{ w_X(\Capx(x)) } \right) .
\end{align*}
Then $\sigmaapx_x \geq \Omega(\sigma_x^\star/(\beta 2^{2z}))$ for all $x\in X$,
and $\sigmaapx_X := \sum_{x \in X} \sigmaapx_x \leq 1 + \alpha k$.
\end{lemma}

Before proving this lemma, we record the following approximate triangle inequality for distances raised to power $z\ge 1$.

\begin{claim} \label{cl:TriangIneq}
For all $x, x', y \in V$ we have
$d^z(x, y) \leq 2^{z-1}\cdot [d^z(x, x') + d^z(x', y)]$.
\end{claim}
\begin{proof}[Proof of \cref{cl:TriangIneq}]
We first use the triangle inequality,
\begin{align*}
  d^z(x,y)
  &\le [d(x, x') + d(x', y)]^z
\intertext{and since $a\mapsto a^z$ is convex (recall $z \ge 1$),
    all $a,b\ge0$ satisfy $(\frac{a+b}{2})^z \le \frac{a^z+b^z}{2}$,
    hence
}
  &\le 2^{z-1} [d^z(x, x') + d^z(x', y)] .
\end{align*}
The claim follows.
\end{proof}

\begin{proof}[Proof of \cref{lemma:sensitivity}]
Given $C^*$, we shorten the notation by setting $\mu:=\NN_{\Capx}$,
and let $\Xapx$ be the weighted set obtained by mapping all points of $X$ by $\mu$.
Formally, $\Xapx := \{ \mu(x) : x \in X \}$
where every $y \in \Xapx$ has weight
$w_{\Xapx}(y) :=  \sum_{x \in X : \mu(x) = y}{w_X(x)} $.
Then obviously
\begin{align*}
  \forall x\in X,
  \quad
  w_X(\Capx(x))
  = \sum_{x'\in \Capx(x)} w_X(x')
  = w_{\Xapx}(\mu(x)) .
\end{align*}

\paragraph{Upper bound on $\sigmaapx_X$.}
Using the above,
\begin{align*}
  \sigmaapx_X
  = \sum_{x \in X}{ \sigmaapx_x }
  = \sum_{x \in X}  w_X(x) \cdot   \left( \frac{d^{z}(x, \mu(x))}{\cost_z(X, \Capx)} + \frac{1}{w_{\Xapx}(\mu(x))} \right)  ,
\end{align*}
and we can bound
\begin{align*}
  \sum_{x \in X}{w_X(x) \cdot \frac{1}{w_{\Xapx}(\mu(x))}}
  = \sum_{y \in \Xapx}{w_{\Xapx}(y) \cdot \frac{1}{w_{\Xapx}(y)}}
  \leq \|\Capx\|_0
  \leq \alpha k,
\end{align*}
and we conclude that
$\sigmaapx_X \leq 1 + \alpha k$,
as required.

\paragraph{Lower bound on $\sigmaapx_x$ (relative to $\sigma_x^\star$).}
Aiming to prove this as an upper bound on $\sigma_x^\star$,
consider for now a fixed $C \in V^k$.
We first establish the following inequality,
that relates the cost of $\Xapx$ to that of $X$.
\begin{align}
  \cost_z(\Xapx, C)
  &= \sum_{y \in \Xapx}{w_{\Xapx}(y) \cdot d^z(y, C)} \nonumber \\
  &= \sum_{x \in X}{w_X(x) \cdot d^z(\mu(x), C)} \nonumber \\
  &\leq 2^{z-1} \sum_{x \in X} w_X(x) \cdot [ d^z(\mu(x), x) + d^z(x, C) ]
    & \text{ by \cref{cl:TriangIneq} }
  \nonumber \\
  &= 2^{z-1}\cdot [ \cost_z(X, \Capx) + \cost_z(X, C) ]
  & \text{ as $\Capx$ is $(\alpha,\beta)$-approximation }
  \nonumber \\
  &\leq 2^{z-1} (\beta + 1) \cdot \cost_z(X, C).
    \label{eq:costXapx}
\end{align}

Now aiming at an upper bound on $\sigma_x^\star$, observe that
\begin{align}
  \frac{d^z(X, C)}{\cost_z(X, C)}
  &\leq 2^{z-1} \cdot \left[ \frac{ d^z(x, \mu(x)) + d^z(\mu(x), C) }{\cost_z(X, C)} \right]
  & \text{ by \cref{cl:TriangIneq} }
    \label{eqn:two_term}
\end{align}
and let us bound each term separately.
For the first term, since $\Capx$ is an $(\alpha, \beta)$-approximation,
\begin{align*}
    \frac{d^z(x, \mu(x))}{\cost_z(X, C)}
    \leq \beta \cdot \frac{d^z(x, \mu(x))}{\cost_z(X, \Capx)}.
\end{align*}
The second term is
\begin{align*}
  \frac{d^z(\mu(x), C)}{\cost_z(X, C)}
  &\leq (\beta + 1) 2^{z-1} \cdot \frac{d^z(\mu(x),C)} {\cost_z(\Xapx, C)}
  & \text{ by~\eqref{eq:costXapx} }
  \\
  &=   (\beta + 1) 2^{z-1} \cdot \frac{d^z(\mu(x),C)} { \sum_{y \in \Xapx}{w_{\Xapx}(y) \cdot d^z(y, C)} }
  \\
  &\leq (\beta+1) 2^{z-1} \cdot \frac{1}{w_{\Xapx}(\mu(x))} .
\end{align*}
Plugging these two bounds into~\eqref{eqn:two_term}, we obtain
\begin{align*}
  \frac{d^z(x, C)}{\cost_z(X, C)}
  \le (\beta+1) 2^{2z-2} \cdot
      \Big[ \frac{d^z(x, \mu(x))}{\cost_z(X, \Capx)}
      + \frac{1}{w_{\Xapx}(\mu(x))} \Big]
  =  (\beta+1) 2^{2z-2} \cdot \frac{\sigmaapx_x}{ w_X(x)} .
\end{align*}
Using the definition in~\eqref{eq:sigmaStar},
we conclude that
$(\beta+1) 2^{2z-2}\cdot \sigmaapx_x \ge \sigma_x^\star$,
which completes the proof of \cref{lemma:sensitivity}.
\end{proof}

\paragraph{Conclusion.}
Our importance sampling algorithm for this type of terminal embedding is listed in \cref{alg:importance_sampling}.
By a direct combination of \cref{lemma:generalized_fl} and \cref{lemma:sensitivity}, we conclude that the algorithm yields a coreset, which is stated formally in \cref{lemma:framework_mult}.
\begin{algorithm}[ht]
    \caption{Coresets for \kzC for $\calF$ with multiplicative distortion}
    \label{alg:importance_sampling}
    \begin{algorithmic}[1]
        \State compute an $(O(1), O(1))$-approximate solution $\Capx$
        for \kzC on $X$ \label{it:ApproxAlg}
        \State for each $x \in X$, let
        $\sigma_x := w_X(x) \cdot \left(\frac{(d(x, \Capx))^z}{\cost_z(X, \Capx)} + \frac{1}{ w_X(\Capx(x)) }\right) $ \label{it:sigmax}
        \Comment{as in \cref{lemma:sensitivity}}
        \State for each $x\in X$, let $p_x := \frac{\sigma_x}{\sum_{y \in X}{\sigma_y}}$
        \State draw $N := O\left(\epsilon^{-2} 2^{2z} k \cdot \left( zk\log{k}\cdot \sdim_{\max}(\calF) + \log{\tfrac1\delta} \right) \right)$
        independent samples from $X$,
        each from the distribution $(p_x:x\in X)$
        \Comment{$\sdim_{\max}$ as in \cref{lemma:generalized_fl}}
        \State let $D$ be the set of samples,
        and assign each $x\in D$ a weight $w_D(x):=\frac{w_X(x)}{p_x N}$
        \State return the weighted set $D$
    \end{algorithmic}
\end{algorithm}

\begin{lemma}
\label{lemma:framework_mult}
Fix $0 < \epsilon, \delta < \frac{1}{2}$, $z\geq 1$, an integer $k \geq 1$,
and a metric space $M(V, d)$.
Given a weighted set $X \subseteq V$
and respective $\calF = \{ f_x : V \to \mathbb{R}_+ \mid x \in X \}$
such that
\begin{align*}
  \forall x \in X, c \in V,
  \qquad
  d(x, c) \leq f_x(c) \leq (1 + \epsilon) \cdot d(x, c),
\end{align*}
\cref{alg:importance_sampling} computes
a weighted set $D \subseteq X$ of size
\begin{align*}
  \|D\|_0 = O\left(\epsilon^{-2} 2^{2z} k \left( zk\log k \cdot \sdim_{\max}(\calF) + \log \tfrac{1}{\delta} \right) \right),
\end{align*}
that with high probability $1 - \delta$ is an $\epsilon$-coreset for \kzC on $X$.
\end{lemma}

The running time of \cref{alg:importance_sampling}
is dominated by the sensitivity estimation, especially line~\ref{it:ApproxAlg} which computes an $(O(1), O(1))$-approximate solution.
In \cref{lemma:alg_2_time} we present efficient implementations of the algorithm,
both in metric settings and in graph settings.

\begin{lemma}
    \label{lemma:alg_2_time}
    \cref{alg:importance_sampling}
    can be implemented
    in time $\tilde{O}(k \|X\|_0)$ if it is given oracle access to the distance $d$,
    and it can be implemented in time $\tilde{O}(|E|)$
    if the input is an edge-weighted graph $G=(V, E)$ and $M$ is its shortest-path metric.
\end{lemma}
\begin{proof}
  The running time is dominated by Step 1 which requires an $(O(1), O(1))$-approximation in both settings.
  For the metric setting where oracle access to $d$ is given,
  \cite{DBLP:journals/ml/MettuP04} gave an $\tilde{O}(k \|X\|_0)$ algorithm
  for both \kMedian ($z = 1$) and \kMeans ($z = 2$), and it has been observed
  to work for general $z$ in a recent work~\cite{HV20}.

  For the graph setting, Thorup~\cite[Theorem 20]{Thorup05}
  gave an $(2, 12 + o(1))$-approximation for graph \kMedian in time $\tilde{O}(|E|)$, such that the input points are \emph{unweighted}.
  Even though not stated in his result,
  we observe that his approach may be easily modified
  to handle weighted inputs as well, and we briefly mention the major changes.
  \begin{itemize}
    \item Thorup's first step~\cite[Algorithm D]{Thorup05}
    is to compute an $(\tilde{O}(\log |V|), O(1))$-approximation $F$ by successive
    uniform independent sampling. This can be naturally modified to sampling
    proportional to the weights of the input points.
    \item Then, the idea is to use the Jain-Vazirani algorithm~\cite{DBLP:journals/jacm/JainV01} on the bipartite graph $F \times X$.
    To make sure the running time is $\tilde{O}(|V|)$,
    the edges of $F \times X$ sub-sampled by picking, for each $x \in X$,
    only $\tilde{O}(1)$ neighbors in $F$.
    This sampling is oblivious to weights, and hence still goes through.
    Let the sampled subgraph be $G'$.
    \item Finally, the Jain-Vazirani algorithm is applied on $G'$ to obtain the
    final $(2, 12 + o(1))$-approximation. However, we still need to
    modify Jain-Vazirani to work with weighted inputs.
    Roughly, Jain-Vazirani algorithm is a primal-dual method,
    so the weights are easily incorporated to the linear program,
    and the primal-dual algorithm is naturally modified so that
    dual variables are increased at a rate that is proportional to their weight in the linear program.
  \end{itemize}

  After obtaining $\Capx$, the remaining steps of \cref{alg:importance_sampling}
  trivially runs in time $\tilde{O}(k \|X\|_0)$ when oracle access to $d$ is given.
  However, for the graph setting, the trivial implementation of Step 2
  which requires to compute $\cost_1(X, \Capx)$ needs to run $\tilde{O}(k)$
  single-source-shortest-paths from points in $\Capx$, and this leads
  to a running time $\tilde{O}(k |V|)$.
  In fact, as observed in~\cite[Observation 1]{Thorup05},
  only one single-source-shortest-path needs to be computed, by running
  Dijkstra's algorithm on a virtual point $x_0$ which
  connects to each point in $\Capx$ to $x_0$ with $0$ weight.

  This completes the proof of \cref{lemma:alg_2_time}.
\end{proof}

\subsection{Coresets via Terminal Embedding with Additive Distortion}
\label{sec:TEadditive}

The second type of embedding has,
in addition to the above $(1+\epsilon)$-multiplicative distortion,
also an additive distortion.
Specifically, we assume the function set $\calF = \calF_S$
is defined with respect to some subset $S \subseteq V$ and satisfies
\begin{align*}
  \forall x \in X, c \in V,
  \quad
  d(x, c) \leq f_x(c) \leq (1 + \epsilon) \cdot d(x, c) + \epsilon \cdot d(x, S).
\end{align*}
The choice of $S$ clearly affects the dimension $\sdim_{\max}(\calF_S)$,
but let us focus now on the effect on the clustering objective,
restricting our attention henceforth only to the case $z=1$ (recalling that $\cost_1 = \cost$).
Suppose we pick $S := \Capx$ where $\Capx$
is an $(\alpha,\beta)$-approximation for \kMedian.
Then even though the additive error for any given $x,y$ might be very large,
it will preserve the \kMedian objective for $X$, because
\begin{align}
  \forall C \in V^k,
  \quad
  \cost(X, C) \leq \cost(\calF, C)
  &\leq (1+\epsilon) \cdot \cost(X, C) + \epsilon \cdot \cost (X, \Capx)
  \nonumber \\
  &\leq (1 + (\beta+1)\epsilon) \cdot \cost(X, C).
    \label{eqn:add_err}
\end{align}
However, this does not immediately imply a coreset for \kMedian,
because we need an analogous bound, but for $D$ instead of $X$
(recall that $D$ is computed by importance sampling with respect to $\calF$).
In particular, using \cref{lemma:generalized_fl} and~\eqref{eqn:add_err}
we get one direction (with high probability)
\begin{align*}
  \forall C\in V^k,
  \quad
  \sum_{x \in D} w_D(x) \cdot f_x(C)
  \geq (1-\epsilon) \cdot \cost(\calF, C)
  \geq (1-\epsilon) \geq \cost(X, C) ,
\end{align*}
however in the other direction we only have
\begin{align*}
  \forall C\in V^k,
  \quad
  \sum_{x \in D} w_D(x) \cdot f_x(C)
  \leq (1 + \epsilon) \cdot \cost(D, X) + \epsilon\cdot \sum_{x \in D}{w_D(x) \cdot d(x, \Capx)},
\end{align*}
where the term $\sum_{x \in D}{w_D(x) \cdot d(x, \Capx)}$
remains to be bounded.

This term $\sum_{x \in D}{w_D(x) \cdot d(x, \Capx)}$
can be viewed as a weak coreset guarantee which preserves the objective
$\cost(X, \cdot)$ on $\Capx$ only.
Fortunately, because $\Capx$ is fixed before the importance sampling,
our algorithm may be interpreted as estimating a fixed sum
\begin{equation*}
  \cost(X, \Capx) = \sum_{x \in X}{w_X(x) \cdot d(x, \Capx)}
\end{equation*}
using independent samples in $D$,
i.e., by the estimator $\sum_{x \in D}{w_D(x) \cdot d(x, \Capx)}$.
And now Hoeffding's inequality shows that for large enough $N$,
this estimator is accurate with high probability.

We present our new algorithm in \cref{alg:importance_sampling_add}, which is largely similar to \cref{alg:importance_sampling},
except for a slightly larger number of samples $N$ and some hidden constants.
Hence, its running time is similar to \cref{alg:importance_sampling},
as stated in \cref{cor:alg_3_time} for completeness.
Its correctness requires new analysis
and is presented in \cref{lemma:framework_add}.

\begin{algorithm}[ht]
\caption{Coresets for \kMedian on $\calF$ with additive distortion}
\label{alg:importance_sampling_add}
\begin{algorithmic}[1]
  \State compute an $(O(1), O(1))$-approximate solution $\Capx$
  for \kMedian on $X$ \label{it:Capx3}
  \State for each $x \in X$, let
  $\sigmaapx_x := w_X(x) \cdot \left(\frac{d(x, \Capx)}{\cost(X, \Capx)} + \frac{1}{ w_X(\Capx(x)) } \right) $ \label{it:sigmaapx3}
  \Comment{as in \cref{lemma:sensitivity}}
  \State for each $x \in X$, let $p_x := \frac{\sigmaapx_x}{\sum_{y \in X}{\sigmaapx_y}}$
  \State draw $N := O\left( \epsilon^{-2} k \left( k\log{k}\cdot \sdim_{\max}(\calF_{\Capx}) + \log \tfrac{1}{\delta} \right)
    + k^2 \log \tfrac{1}{\delta} \right) $
  independent samples from $X$, each from the distribution $(p_x : x \in X)$
  \label{it:N3}
  \Comment{$\sdim_{\max}$ as in \cref{lemma:generalized_fl},
    and $\calF_{\Capx}$ as in~\eqref{eqn:add_statement}
  }
  \State for each $x$ in the sample $D$
  assign weight $w_D(x) := \frac{w_X(x)}{p_x N}$
  \State return the weighted set $D$
\end{algorithmic}
\end{algorithm}

\begin{corollary}
    \label{cor:alg_3_time}
    \cref{alg:importance_sampling_add}
    can be implemented
    in time $\tilde{O}(k \|X\|_0)$ if it is given oracle access to the distance $d$,
    and in time $\tilde{O}(|V| + |E|)$
    if the input is an edge-weighted graph $G=(V, E)$ and $M$ is its shortest-path metric.
\end{corollary}

\begin{lemma}
\label{lemma:framework_add}
Fix $0 < \epsilon, \delta <\frac{1}{2}$, an integer $k \geq 1$,
and a metric space $M(V, d)$.
Given a weighted set $X \subseteq V$,
and an $(O(1), O(1))$-approximate solution $\Capx\in V^k$ for \kMedian on $X$,
suppose $\calF_\Capx = \{ f_x : V \to \mathbb{R}_+ \mid x \in X \}$ satisfies
\begin{align}
\label{eqn:add_statement}
  \forall x \in X, c \in V,
  \quad
  d(x, c) \leq f_x(c) \leq (1 + \epsilon) \cdot d(x, c) + \epsilon \cdot d(x, \Capx) ;
\end{align}
then \cref{alg:importance_sampling_add}
computes a weighted set $D \subseteq X$ of size
\begin{align*}
  \|D\|_0 = O\left(\epsilon^{-2} k \left( k\log{k}\cdot \sdim_{\max}(\calF_{\Capx}) + \log\tfrac{1}{\delta} \right)
  + k^2\log \tfrac{1}{\delta} \right),
\end{align*}
that with high probability $1 - \delta$
is an $\epsilon$-coreset for \kMedian on $X$.
\end{lemma}

\begin{proof}
Suppose $\Capx\in V^k$
is an $(\alpha, \beta)$-approximate solution for $\alpha,\beta=O(1)$.
Observe that~\eqref{eqn:add_statement} implies~\eqref{eqn:add_err},
and write $\calF = \calF_{\Capx}$ for brevity.

\paragraph{Sensitivity Analysis.}
We would like to employ \cref{lemma:generalized_fl}.
Observe that $\sigmaapx_x$ in \cref{alg:importance_sampling_add}
is the same, up to hidden constants,
as in \cref{alg:importance_sampling},
hence the upper bound $\sigmaapx_X \leq 1 +\alpha k$
follows immediately from \cref{lemma:sensitivity}.
We also need to prove that
$\sigmaapx_x \geq \Omega(\sigma^\calF_x)$ for all $x \in X$,
where
$\sigma^\calF_x = \max_{C \in V^k}{\frac{w_X(x) \cdot f_x(C)}{\cost(\calF, C)}}$.
Once again, we aim to prove this as an upper bound on $\sigma^\calF_x$.

Fix $x\in X$,
and let $\Cmax\in V^k$ be a maximizer in the definition of $\sigma^\calF_x$
(which clearly depends on $x$).
Then
\begin{align*}
  \sigma^\calF_x 
  &= \frac{w_X(x) \cdot f_x(\Cmax)}{\cost(\calF, \Cmax)} \\
  &\leq \frac{ w_X(x)\cdot [(1+\epsilon)\cdot d(x,\Cmax) + \epsilon\cdot d(x,\Capx)] }{ \cost(X, \Cmax) }
  & \text{by~\eqref{eqn:add_statement} and~\eqref{eqn:add_err}}
  \\
  &\leq (1+\epsilon)\cdot \sigma^\star_x + \epsilon\cdot \frac{ w_X(x)\cdot d(x,\Capx) }{ \cost(X, \Cmax) }
  & \text{as defined in~\eqref{eq:sigmaStar}}
  \\
  &\leq (1+\epsilon)\cdot \sigma^\star_x + \beta\epsilon\cdot \frac{ w_X(x)\cdot d(x,\Capx) }{ \cost(X, \Capx) }
  & \text{as $\Capx$ is $(\alpha,\beta)$-approximation}
  \\
  &\leq (1+\epsilon)\cdot \sigma^\star_x + \beta\epsilon\cdot \sigmaapx_x .
  & \text{as defined in line~\ref{it:sigmaapx3}}
\end{align*}
Combining this with our bound
$\sigma_x^\star \leq O(\beta)\cdot \sigmaapx_x$
from \cref{lemma:sensitivity} (recall $z=1$),
we conclude that
$\sigma^\calF_x \leq O(\beta)\cdot \sigmaapx$.

\paragraph{Overall Error Bound.}
Recall our goal is to prove that with probably at least $1 - \delta$,
the output $D$ is a coreset, i.e.,
\begin{align} \label{eq:add_goal}
  \forall C\in V^k,
  \quad
  \cost(D, C) \in (1\pm O(\beta \epsilon)) \cdot \cost(X, C).
\end{align}
Applying \cref{lemma:generalized_fl}
with our choice of $N$ in line~\ref{it:N3} of the algorithm,
we know that with probability at least $1 - \delta/2$,
\begin{align}
  \forall C \in V^k,\quad
  \sum_{x \in D}{ w_D(x) \cdot f_x(C) } \in (1 \pm \epsilon)
  \cdot \cost(\calF, C)
  \label{eqn:coreset}
\end{align}
We claim, and will prove shortly, that with probability at least $1-\delta/2$,
\begin{align} \label{eq:DeviationD}
  \sum_{x \in D}{w_D(x) \cdot d(x, \Capx)} \leq
  2 \cdot \cost(X, \Capx) .
\end{align}
Using this claim, we complete the proof as follows.
By a union bound, with probability at least $1-\delta$,
both~\eqref{eqn:coreset} and~\eqref{eq:DeviationD} hold.
In this case, for all $C \in V^k$,
one direction of~\eqref{eq:add_goal} follows easily
\begin{align*}
  \cost(D, C)
&\leq \sum_{x \in D} {w_D(x) \cdot f_x(C)}
  &\text{by~\eqref{eqn:add_statement}}
  \\
  &\leq (1 + \epsilon) \cdot \cost(\calF, C)
  &\text{by~\eqref{eqn:coreset}}
  \\
  &\leq (1 + O((\beta\epsilon)) \cdot \cost(X, C) .
  &\text{by~\eqref{eqn:add_err}}
\end{align*}
For the other direction of~\eqref{eq:add_goal},
which crucially rely on~\eqref{eq:DeviationD},
we have
\begin{align*}
  \cost(X, C)
  &\leq \cost(\calF, C)
  &\text{by~\eqref{eqn:add_err}}
  \\
  &\leq \frac{1}{1 - \epsilon} \cdot \sum_{x \in D}{w_D(x) \cdot f_x(C)}
  &\text{by~\eqref{eqn:coreset}}
  \\
  &\leq \frac{1 + \epsilon}{1 - \epsilon} \cdot \cost(D, C)
    + \frac{\epsilon}{1 - \epsilon} \cdot \sum_{x \in D}{w_D(x) \cdot d(x, \Capx)}
  &\text{by~\eqref{eqn:add_statement}}
  \\
  &\leq \frac{1 + \epsilon}{1 - \epsilon} \cdot \cost(D, C)
    + \frac{2\epsilon}{1 - \epsilon} \cdot \cost(X, \Capx) ,
  & \text{by~\eqref{eq:DeviationD}}
\end{align*}
and finally using that $\Capx$ is $(\alpha,\beta)$-approximation
and some rearrangement, we get that
$\cost(X, C) \leq (1+O(\beta\epsilon)) \cost(D, C)$.

It remains to prove our claim,
i.e., that~\eqref{eq:DeviationD} holds with high probability.
This follows by a straightforward application of Hoeffding's Inequality.
To see this, define for each $1 \leq i \leq N$ the random variable
$Y_i := \frac{w_X(x) \cdot d(x, \Capx)}{p_x}$,
where $x$ is the $i$-th sample in line~\ref{it:N3},
and let
$Y := \frac{1}{N} \sum_{i = 1}^{N}{Y_i}$.
Then
\[
  Y = \sum_{x \in D}{w_D(x) \cdot d(x, \Capx)},
\]
and its expectation is
$\E[Y] = \E[Y_1] = \sum_{x \in X}{w_X(x) \cdot d(x, \Capx)} = \cost(X, \Capx)$.

Now observe that the random variables $Y_i$ are independent,
and use \cref{lemma:sensitivity} to bound each of them by
\begin{align*}
  0 \leq Y_i
= \frac{ w_X(x) \cdot d(x, \Capx) }{ \sigmaapx_x/\sigmaapx_X }
  \leq (1+\alpha k) \cdot \cost(x, \Capx)
  = (1+\alpha k) \E[Y] .
\end{align*}
Hence, by Hoeffding's Inequality
\begin{align*}
  \forall t>0,
  \quad
  \Pr\left[ Y - \E[Y] > t \right]
  \leq \exp\left( - \frac{2N t^2 }{ ((1+\alpha k) \E[Y])^2 } \right)
\end{align*}
and for $t=\E[Y]$
and a suitable $N \ge \Omega(\alpha^2 k^2 \log \tfrac{1}{\delta})$,
we conclude that
$\Pr\big[Y > 2\E[Y] \big] \leq \delta/2$.
This proves the claim
and completes the proof of \cref{lemma:framework_add}.
\end{proof}

 \section{Coresets}
\label{sec:coresets}

\ifsubmission
As discussed in \cref{sec:intro} (and formally stated in \cref{sec:framework}),
efficient coreset constructions follow from terminal embeddings $\calF$ with low distortion and dimension.
\else
We now apply the framework developed in \cref{sec:framework}
to design coresets of size independent of $X$ for various settings,
including excluded-minor graphs (in \cref{sec:ExcludedMinor}),
high-dimensional Euclidean spaces (in \cref{sec:Euclidean}),
and graphs with bounded highway dimension (in \cref{sec:hw}).
Our workhorse will be \cref{lemma:framework_mult} and \cref{lemma:framework_add},
which effectively translate a terminal embedding $\calF$
with low distortion on $X \times V$
and low shattering dimension $\sdim_{\max}$
into an efficient algorithm to construct
a coreset whose size is linear in $\sdim_{\max}(\calF)$.

\fi
We therefore turn our attention to designing various terminal embeddings.
For excluded-minor graphs, we design a terminal embedding $\calF$
with multiplicative distortion $1 + \epsilon$ of the distances,
and dimension $\sdim_{\max}(\calF) = O(\poly(k/ \epsilon)\cdot \log{\|X\|_0})$.
For Euclidean spaces, we employ a known terminal embedding with similar guarantees. 
In both settings, even though the shattering dimension depends on ${\|X\|_0}$,
it still implies coresets of size independent of $X$
by our iterative size reduction (\cref{thm:ite_size_reduct}).
We thus obtain the first coreset (of size independent of $X$ and $V$) 
for excluded-minor graphs (\cref{cor:coreset_mf}),
and a simpler state-of-the-art coreset
for Euclidean spaces (\cref{cor:coreset_euclidean}).

We also design a terminal embedding for graphs with bounded highway dimension
(formally defined in \cref{sec:hw}). 
This embedding has an additive distortion (on top of the multiplicative one),
but its shattering dimension is independent of $X$,
hence the iterative size reduction is not required. 
We thus obtain the first coreset (of size independent of $X$ and $V$)
for graphs with bounded highway dimension (\cref{cor:coreset_hw}).

\subsection{Excluded-minor Graphs}
\label{sec:ExcludedMinor}

Our terminal embedding for excluded-minor graphs is stated in the next lemma.
Previously, the shattering dimension of the shortest-path metric
of graphs excluding a fixed graph $H_0$ as a minor was studied only for unit point weight,
for which Bousquet and Thomass{\'{e}}~\cite{DBLP:journals/dm/BousquetT15}
proved that $\calF = \{ d(x, \cdot) \mid x \in X\}$
has shattering dimension $\sdim(\calF) = O(|H_0|)$. 
For arbitrary point weight, i.e., $\sdim_{\max}(\calF)$,
it is still open to get a bound that depends only on $|H_0|$,
although the special case of bounded treewidth was recently resolved,
as Baker et al.~\cite{coreset_tw},
proved that $\sdim_{\max}(\calF) = O(\tw(G))$
where $\tw(G)$ denotes the treewidth of the graph $G$. 
Note that both of these results use no distortion of the distances, 
i.e., they bound $\calF = \{ d(x, \cdot) \mid x \in X \}$.
Our terminal embedding handles the most general setting 
of excluded-minor graphs and arbitrary point weight,
although it bypasses the open question 
by allowing a small distortion and dependence on $X$.

\begin{lemma}[Terminal Embedding for Excluded-minor Graphs]
\label{lemma:mf_sdim}
For every edge-weighted graph $G = (V, E)$ that excludes some fixed minor
and whose shortest-path metric is denoted as $M = (V, d)$,
and for every weighted set $X \subseteq V$,
there exists a set of functions
$\calF := \{ f_x : V \to \mathbb{R}_+ \mid x \in X \}$
such that
\[
  \forall x \in X, c \in V,
  \qquad
  d(x, c) \leq f_x(c) \leq (1 + \epsilon) \cdot d(x, c),
\]
and $\sdim_{\max}(\calF) = \tilde{O}(\epsilon^{-2}) \cdot  \log{\|X\|_0}$.
\end{lemma}

Let us present now an overview of the proof of \cref{lemma:mf_sdim},
deferring the full details to \cref{sec:proof_mf}. 
Our starting point is the following approach,
which was developed in~\cite{coreset_tw}
for bounded-treewidth graphs.
(The main purpose is to explain how vertex separators are used as portals
to bound the shattering dimension, 
but unfortunately additional technical details are needed.) 
The first step in this approach reduces
the task of bounding the shattering dimension
to counting how many distinct permutations of $X$ one can obtain
by ordering the points of $X$ according to their distance from a point $c$,
when ranging over all $c\in V$.
An additional argument uses the bounded treewidth 
to reduce the range of $c$ from all of $V$ to a subset $\hatV\subset V$,
that is separated from $X$ by a vertex-cut $P\subset V$ of size $|\hatP|=O(1)$. 
This means that every path, including the shortest-path, 
between every $x\in X$ and every $c\in \hatV$ must pass through $\hatP$, 
therefore
\[
  d(x,c) = \minn{ d(x,p) + d(p,c): p\in \hatP },
\]
and the possible orderings of $X$ are completely determined by these values. 
The key idea now is to replace the hard-to-control range of $c\in\hatV$
with a richer but easier range of $|\hatP|=O(1)$ real variables.
Indeed, each $d(x,\cdot)$ is captured by a \emph{min-linear function}, 
which means a function of the form $\min_{i}{a_i y_i + b_i}$
with real variables $\set{y_i}$ that represent $\set{d(p,c)}_{p\in \hatP}$ 
and fixed coefficients $\set{a_i,b_i}$. 
Therefore, each $d(x,\cdot)$
is captured by a min-linear function $g_x: \RR^{|\hatP|} \to \RR_+$,
and these functions are all defined on the same $|\hatP|=O(1)$ real variables. 
In this representation, it is easy to handle the point weight $v : X \to \RR_+$
(to scale all distances from $x$), 
because each resulting function $v(x)\cdot g_x$ is still min-linear.
Finally, the number of orderings of
the set $\set{g_x}_{x\in X}$ of min-linear functions,
is counted using the arrangement number for hyperplanes,
which is a well-studied quantity in computational geometry.

To extend this approach to excluded-minor graphs (or even planar graphs),
which do not admit small vertex separators,
we have to replace vertex separators with shortest-path separators~\cite{DBLP:journals/jacm/Thorup04, DBLP:conf/podc/AbrahamG06}.
In particular,
  we use these separator theorem to partition the whole graph into a few parts,
  such that each part is separated from the graph by only a few shortest paths,
  see \cref{lemma:planar_sep} for planar graphs
  (which is a variant of a result known from~\cite{DBLP:conf/soda/EisenstatKM14})
  and \cref{lemma:mf_sep} for excluded-minor graphs.
However, the immediate obstacle is that while these separators consist of a few paths,
their total size is unbounded (with respect to $X$), 
which breaks the above approach
because each min-linear function has too many variables. 
A standard technique to address this size issue 
is to discretize the path separator into \emph{portals}, 
and reroute through them a shortest-path from each $x\in X$ to each $c\in V$.
This step distorts the distances, 
and to keep the distortion bounded multiplicatively by $1+\epsilon$,
one usually finds inside each separating shortest-path $l$,
a set of portals $P_l\subset l$ whose spacing is at most $\epsilon\cdot d(x,c)$.
However, $d(x,c)$ could be very small compared to the entire path $l$, 
hence we cannot control the number of portals (even for one path $l$).

\paragraph{Vertex-dependent Portals}
In fact, all we need is to represent the relative ordering of
$\{ d(x, \cdot) : x\in X \}$ using a set of \emph{min-linear functions} over a few real variables, and these variables do not have to be the distance to \emph{fixed portals} on the separating shortest paths.
(Recall this description is eventually used by
  the arrangement number of hyperplanes to count orderings of $X$.) 
To achieve this, we first define \emph{vertex-dependent} portals $P^{l}_c$
with respect to a separating shortest path $l$ \emph{and} a vertex $c\in V$
(notice this includes also $P^{l}_x$ for $x \in X$).
and then a shortest path from $x \in X$ to $c \in V$ passing through $l$
is rerouted through portals $P^l_x \cup P^l_c$, as follows.
First, since $l$ is itself a shortest path,
$d(x, c) = \min_{u_1, u_2 \in l}\{d(x, u_1) + d(u_1, u_2) + d(u_2, c)\}$.
Observe that $d(u_1, u_2)$ is already linear,
because one real variable can ``capture'' a location in $l$,
hence we only need to approximate $d(x, u_1)$ and $d(c, u_2)$. 
To do so, we approximate the distances from $c$ to every vertex on the path $l$,
i.e., $\set{d(c,u)}_{u\in l}$,
using only the distances from $c$ to its portal set $P^l_c$,
i.e., $\set{d(c,p)}_{p\in P^l_c}$. 
Moreover, between successive portals this approximate distance is a linear function,
and it actually suffices to use $|P^l_c| = \poly(1/\epsilon)$ portals,
which means that $d(c,u)$ can be represented as a \emph{piece-wise linear} function in $\poly(1/\epsilon)$ real variables.

Note that the above approach ends up with 
the minimum of piece-wise linear (rather than linear) functions,
which creates extra difficulty.
In particular, we care about the relative ordering of $\{ d(x, \cdot) : x \in X \}$
over all $c \in V$,
and to evaluate $d(x, c)$ we need the pieces that $c$ and $x$ generate,
i.e., information about $P^l_c\cup P^l_x$. 
Since the number of $c \in V$ is unbounded, we need to ``guess'' the
structure of $P^l_c$, specifically the ordering between the portals
in $P^l_c$ and those in $P^l_x$. 
Fortunately, since every $|P^l_c| \leq \poly(1/\epsilon)$, 
such a ``guess'' is still affordable,
and this would prove \cref{lemma:mf_sdim}.

\begin{corollary}[Coresets for Excluded-Minor Graphs]
\label{cor:coreset_mf}
For every edge-weighted graph $G = (V, E)$
that excludes a fixed minor,
every $0 < \epsilon, \delta < 1/2$ and integer $k \geq 1$,
\kMedian of every weighted set $X \subseteq V$
(with respect to the shortest path metric of $G$)
admits an $\epsilon$-coreset of size
$\tilde{O}(\epsilon^{-4} k^2 \log{\frac{1}{\delta}})$.
Furthermore, such a coreset can be computed in time $\tilde{O}(|E|)$
with success probability $1 - \delta$. 
\end{corollary}

\begin{proof}
By combining \cref{lemma:framework_mult}, \cref{lemma:alg_2_time}
with our terminal embedding from \cref{lemma:mf_sdim},
we obtain an efficient algorithm for constructing a coreset
of size $\tilde{O}(\epsilon^{-4} k^2 \log{\|X\|_0})$.
This size can be reduced to the claimed size (and running time) 
using the iterative size reduction of \cref{thm:ite_size_reduct}.
\end{proof}

\begin{remark}
This result partly extends to \kzC for all $z\ge 1$. 
The importance sampling algorithm and its analysis are immediate,
and in particular imply the existence of a coreset of size
$\tilde{O}(\epsilon^{-4} k^2 \log{\frac{1}{\delta}})$.
However we rely on known algorithm for $z=1$ in the step of
computing an approximate clustering (needed to compute sampling probabilities). 
\end{remark}

\subsection{Proof of \cref{lemma:mf_sdim}}
\label{sec:proof_mf}
For the sake of presentation, we start with proving the planar case,
since this already requires most of our new technical ideas.
The statement of terminal embedding for planar graphs is as follows,
and how the proof can be modified to work for the minor-excluded case
is discussed in \cref{sec:planar_to_mf}.

\begin{lemma}[Terminal Embedding for Planar Graphs]
    \label{lemma:planar_sdim}
    For every edge-weighted planar graph $G = (V, E)$ whose shortest path metric is denoted as $M = (V, d)$
    and every weighted set $X \subseteq V$,
    there exists a set of functions $\calF = \calF_X := \{ f_x : V \to \mathbb{R}_+ \mid x \in X \}$ such that
    for every $x \in X$, and $c \in V$, $f_x(c) \in (1 \pm \epsilon) \cdot d(x, c)$, and $\sdim_{\max}(\calF) = \widetilde{O}(\epsilon^{-2}) \log{\|X\|_0}$.
\end{lemma}

By definition, $\sdim_{\max}(\calF) = \max_{v : X \to \mathbb{R}_+}(\calF_v)$,
so it suffices to bound $\sdim(\calF_v)$ for every $v$.
Also, by the definition of $\sdim$, it suffices to prove for every $\calH \subseteq \calF_v$ with $|\calH| \geq 2$,
\begin{align*}
    \left| \{ B_{\calH}(c, r) : c \in V, r\geq 0 \} \right|
    \leq \poly(\|X\|_0) \cdot |\calH|^{\tilde{O}(\epsilon^{-2})\log{\|X\|_0}}.
\end{align*}
Hence, we fix some $v : X \to \mathbb{R}_+$ and $\calH \subseteq \calF_v$
with $|\calH| \geq 2$ throughout the proof.

\paragraph{General Reduction: Counting Relative Orderings}
For $\calH \subseteq \calF$ and $c\in V$, let $\sigma^{\calH}_c$ be the permutation of $\calH$ ordered by
$ v(x) \cdot f_x(c) $ in non-decreasing order and ties are broken arbitrarily.
Then for a fixed $c \in V$ and very $r \geq 0$,
the subset $B_\calH(c, r) \subseteq \calH$ is exactly the subset defined by some prefix of $\sigma_c^\calH$.
Hence,
\begin{align*}
    \left| \{B_{\calH}(c, r) : c\in V, r \geq 0\} \right|
    \leq |\calH| \cdot \left| \{ \sigma^{\calH}_c : c \in V \} \right|.
\end{align*}
Therefore, it suffices to show
\begin{align*}
    \left| \{ \sigma_c^{\calH} : c \in V \} \right| \leq \poly(\|X\|_0) \cdot  |\calH|^{\tilde{O}(\epsilon^{-2}) \log{\|X\|_0}}.
\end{align*}
Hence, this reduces the task of bounding of shattering dimension to counting
the number of relative orderings of $\{v(x) \cdot f_x(c) \mid x\in X\}$.

Next, we use the following structural lemma for planar graphs
to break the graph into few parts of simple structure,
so we can bound the number of permutations for $c$ coming from each part.
A variant of this lemma has been proved in~\cite{DBLP:conf/soda/EisenstatKM14},
where the key idea is to use the \emph{interdigitating} trees.
For completeness,
  we give a full proof of this lemma in \cref{sec:proof_planar_sep}.
\begin{lemma}[Structural Property of Planar Graphs, see also~\cite{DBLP:conf/soda/EisenstatKM14}]
    \label{lemma:planar_sep}
    For every edge-weighted planar graph $G = (V, E)$ and subset $S \subseteq V$,
    $V$ can be broken into parts $\Pi := \{ V_i \}_i$
    with $|\Pi| = \mathrm{poly}(|S|)$ and $\bigcup_{i}{V_i} = V$, such that
    for every $V_i \in \Pi$,
    \begin{enumerate}
        \item $|S \cap V_i| = O(1)$,
        \item there exists a collection of shortest paths $\calP_i$ in $G$ with $|\calP_i| = O(1)$ and
        removing the vertices of all paths in $\calP_i$ disconnects $V_i$ from $V\setminus V_i$ (points in $V_i$ are possibly removed).
    \end{enumerate}
    Furthermore, such $\Pi$ and the corresponding shortest paths $\calP_i$
    for $V_i \in \Pi$ can be computed in $\tilde{O}(|V|)$ time\footnote{This lemma is used only in the analysis in this section,
    but the running time is relevant when this lemma is used again in
    \cref{sec:application}.}.
\end{lemma}

Applying \cref{lemma:planar_sep} with $S = X$ (noting that $S$ is an unweighted set),
we obtain $\Pi = \{V_i\}_i$ with $|\Pi| = \mathrm{poly}(\|X\|_0)$,
such that each part $V_i \in \Pi$ is separated by $O(1)$ shortest paths $\calP_i$. Then
\begin{align*}
    \left| \{ \sigma_c^{\calH} : c \in V \} \right|
    \leq \sum_{V_i \in \Pi}{ \left| \{\sigma_c^{\calH} : c \in V_i \}\right| }.
\end{align*}
Hence it suffices to show for every $V_i \in \Pi$, it holds that
\begin{align}
    \left| \{ \sigma_c^{\calH} : c \in V_i \} \right|
    \leq |\calH|^{\tilde{O}(\epsilon^{-2}) \log{\|X\|_0}}.
    \label{eqn:sigma}
\end{align}
Since $\bigcup_{i}{V_i} = V$, it suffices to define functions $f_x(\cdot)$ for $c \in V_i$ for every $i$ independently.
Therefore, we fix $V_i \in \Pi$ throughout the proof.
In the following, our proof proceeds in three parts.
The first defines functions $f_x(\cdot)$ on $V_i$,
the second analyzes the distortion of $f_x$'s,
and the final part analyzes the shattering dimension.

\paragraph{Part I: Definition of $f_x$ on $V_i$}
By \cref{lemma:planar_sep} we know $|V_i \cap X|= O(1)$.
Hence, the ``simple'' case is when $x \in V_i \cap T$,
for which we define $f_x(\cdot) := d(x, \cdot)$.

Otherwise, $x \in X\setminus V_i$. Write $\calP_i := \{ P_j \}_j$.
Since $P_j$'s are shortest paths in $G$,
and removing $\calP_i$ from $G$ disconnects $V_i$ from $V \setminus V_i$,
we have the following fact.
\begin{fact}
    \label{fact:sp_cross}
    For $c \in V_i$ and $x \in X \setminus V_i$, there exists $P_j \in \calP_i$ and $c', x' \in P_j$,
    such that $d(c, x) = d(c, c') + d(c', x') + d(x', x)$.
\end{fact}
Let $d_j(c, x)$ be the length of the shortest path from $c$ to $x$ that uses \emph{at least one} point in $P_j$.
For each $P_j \in \calP_i$, we will define $f_x^{j} : V_i \to \mathbb{R}_+$,
such that $f_x^{j}(c)$ is within $(1 \pm \epsilon) \cdot d_j(c, x)$, and
let
\begin{align*}
    f_x(c) := \min_{P_j \in \calP_i}{f_x^{j}(c)}, \qquad \forall c \in V_i.
\end{align*}
Hence, by \cref{fact:sp_cross}, the guarantee that $f_x^{j}(c) \in (1\pm \epsilon) \cdot d_j(c, x)$
implies $f_x(c) \in (1\pm \epsilon) \cdot d(x, c)$,
as desired.
Hence we focus on defining $f_x^{j}$ in the following.

\paragraph{Defining $f_x^{j} : V_i \to \mathbb{R}_+$}
Suppose we fix some $P_j \in \calP_i$, and we will define $f_x^{j}(c)$, for $c \in V_i$.
By \cref{fact:sp_cross} and the optimality of shortest paths, we have
\begin{align*}
    d_j(x, c) = \min_{c', x' \in P_j}\{d(c, c') + d(c', x') + d(x', x) \}.
\end{align*}
For every $y \in V$, we will define
$l_y^{j} : P_j \to \mathbb{R}_+$
such that $l_y^{j}(y') \in (1 \pm \epsilon) \cdot d(y, y')$ for every $y' \in P_j$.
Then, we let
\begin{align*}
    f_x^{j}(c) := \min_{c', x' \in P_j}\{l_c^{j}(c') + d(c', x') + l_x^{j}(x')\},
\end{align*}
and this would imply $f_x^{j}(c) \in (1\pm \epsilon) \cdot d_j(x, c)$.
So it remains to define $l_y^j : P_j \to \mathbb{R}_+$ for every $y \in V$.

\paragraph{Defining $l_y^{j} : P_j \to \mathbb{R}_+$}
Fix $y\in V$ and we will define $l_y^{j}(y')$ for every $y' \in P_j$.
Pick $h_y \in P_j$ that satisfies $d(y, h_y) = d(y, P_j)$.
Since $P_j$ is a shortest path, we interpret $P_j$ as a segment in the real line.
In particular, we let the two end points of $P_j$ be $0$ and $1$, and $P_j$ is a (discrete) subset of $[0, 1]$.

Define $a, b\in P_j$ such that $a \leq h_y \leq b$
are the two \emph{furthest} points on the two sides of $h$ on $P_j$ that satisfy
$d(h_y, a) \leq \frac{d(y, h_y)}{\epsilon}$ and $d(h_y, b) \leq \frac{d(y, h_y)}{\epsilon}$.
Then construct a sequence of points $a = q_1 \leq q_2 \ldots$ in the following way.
For $t = 1, 2,\ldots$, if there exists $u \in (q_t, 1] \cap P_j$
such that $d(q_t, u) > \epsilon \cdot d(y, h_y)$,
then let $q_{t+1}$ be the smallest such $u$;
if such $u$ does not exist, then
let $q_{t+1} := b$ and terminate.
Essentially, this breaks $P_j$ into segments of length $\epsilon \cdot d(y, h_y)$,
except that the last one that ends with $b$ may be shorter.
Denote this sequence as $Q_y := (q_1 = a, \ldots, q_m = b)$.
\begin{claim}
    \label{claim:m_ub}
    For every $y \in V$, $|Q_y| = O(\epsilon^{-2})$.
\end{claim}
\begin{proof}
    By the definition of $Q_y$, for $1 \leq t \leq m - 2$, $d(q_{t}, q_{t+1}) > \epsilon \cdot d(y, h_y) $.
    On the other hand, by the definition of $a$ and $b$, $d(q_1, q_m) = d(a, b) \leq O(\frac{d(y, h_y)}{\epsilon})$.
    Therefore, $|Q_y| \leq O(\epsilon^{-2})$, as desired.
\end{proof}

\paragraph{Definition of $f_x$ on $V_i$: Recap}
Define
\begin{align}
    l_y^{j}(y') :=
    \begin{cases}
        d(h_y, y') & \text{ if } y' < a = q_1 \text{ or } y' > b = q_m \\
        d(y, q_t) & \text{ if } q_t \leq y' < q_{t+1}, 1 \leq t < m  \\
        d(y, q_m) & \text{ if } y' = b = q_m
    \end{cases}
    \label{eqn:def_l}
\end{align}
where $h_y \in P_j$, $Q_y = \{q_t\}_t \subset P_j$.
To recap,
\begin{itemize}
    \item if $x \in X \cap V_i$, then $f_x(c) := d(x, c)$;
    \item otherwise $x \in X \setminus V_i$,
    $f_x(c) := \min_{P_j \in \calP_i}{f_x^{j}(c)}$, where
        \begin{align}
            f_x^{j}(c) := \min_{c', x' \in P_j}\{
                    l_c^{j}(c') + d(c', x') + l_x^{j}(x')
                \}.
                \label{eqn:def_fxj}
        \end{align}
\end{itemize}
Finally,
\begin{align}
    f_x(c) := \min_{P_j \in \calP_i}{f_x^{j}(c)}, \qquad \forall c \in V_i.
    \label{eqn:def_fx}
\end{align}

\paragraph{Part II: Distortion Analysis}
The distortion of $l$'s is analyzed in the following \cref{lemma:l_error}, and the distortion for $f_x$ follows immediately from the above definitions.
\begin{lemma}
    \label{lemma:l_error}
    For every $P_j \in \calP_i$, $y \in V$, $y' \in P_j$,
    $l_y^{j}(y') \in (1 \pm \epsilon) \cdot d(y, y')$.
\end{lemma}
\begin{proof}
    If $y' = q_m = b$, by definition $l_y^{j}(y') = d(y, q_m) = d(y, y')$.
    Then consider the case when $y' < a = q_1$ or $y'> b = q_m$.
    \begin{align*}
        l_y^{j}(y')
        &= d(h_y, y') \\
        &\in d(y', y) \pm d(y, h_y) \\
        &\in d(y', y) \pm \epsilon \cdot d(y', h_y),
    \end{align*}
    where the last inequality follows from $d(y', h_y) > \frac{d(y, h_y)}{\epsilon}$.
    This implies $d(y, y') \in (1 \pm \epsilon) \cdot l_y^{j}(y')$.

    Otherwise, $q_t \leq y' < q_{t+1}$ for some $1\leq t < m$.
    By the definition of $q_t$'s and the definition of $h_y$,
    \begin{align*}
        d(y, y')
        &\in d(y, q_t) \pm d(q_t, y') \\
        &\in d(y, q_t) \pm \epsilon \cdot d(y, h_y) \\
        &\in d(y, q_t) \pm \epsilon \cdot d(y, y') \\
        &\in l_y^{j}(y') \pm \epsilon \cdot d(y, y'),
    \end{align*}
    which implies $l_y^{j}(y') \in (1 \pm \epsilon) \cdot d(y, y')$.
    This finishes the proof of \cref{lemma:l_error}.
\end{proof}

\paragraph{Part III: Shattering Dimension Analysis}
Recall that we fixed $v : X \to \mathbb{R}_+$ and $\calH \subseteq \calF_v$ with $|\calH| \geq 2$. Now we show
\begin{align}
    \label{eqn:sigma_vi_goal}
    \left| \{\sigma_c^{\calH} : c \in V_i \} \right| \leq |\calH|^{\tilde{O}(\epsilon^{-2})\log{\|X\|_0}}.
\end{align}
Let $H := \{ x : v(x) \cdot f_x \in \calH \}$, so $|H| = |\calH|$.
Recall that $|V_i \cap X| = O(1)$ by \cref{lemma:planar_sep},
so $|V_i \cap H| = O(1)$.
Hence, if we could show
\begin{align*}
    \left| \{ \sigma_c^{\calH} : c \in V_i \} \right| \leq N(|H|)
\end{align*}
for $\calH$ such that $H \cap V_i = \emptyset$,
then for general $\calH$,
\begin{align*}
    \left| \{ \sigma_c^{\calH} : c \in V_i \} \right| \leq N(|H| - |V_i \cap H|) \cdot |H|^{O(|V_i \cap H|)} \leq N(|H|) \cdot |H|^{O(1)}.
\end{align*}
Therefore, it suffices to show~\eqref{eqn:sigma_vi_goal} under the assumption that
$H \cap V_i = \emptyset$.

In the following, we will further break $V_i$ into $|H|^{\tilde{O}(\epsilon^{-2})}$ parts,
such that for each part $V'$, $f_x$ on $V'$ may be alternatively represented as a \emph{min-linear} function.
\begin{lemma}
    \label{lemma:min_linear_partition}
    Let $u = |\calP_i|$.
    There exists a partition $\Gamma$ of $V_i$, such that the following holds.
    \begin{enumerate}
        \item $|\Gamma| \leq |H|^{\tilde{O}(\epsilon^{-2})\cdot u}$.
        \item \label{itm:gx} $\forall V' \in \Gamma$, $\forall x \in H$,
        there exists $g_x : \mathbb{R}^s \to \mathbb{R}_+$ where $s = O(\epsilon^{-2})$,
        such that $g_x$ is a minimum of $O(\epsilon^{-4} u)$ linear functions on $\mathbb{R}^s$,
        and for every $c \in V'$, there exists $y \in \mathbb{R}^s$ that satisfies $f_x(c) = g_x(y)$.
    \end{enumerate}
\end{lemma}
\begin{proof}
    Before we actually prove the lemma, we need to examine $f_x^j(c)$ and $l_y^{j}$ more closely. Suppose some $P_j \in \calP_i$ is fixed.
    Recall that for $y \in V, y' \in P_j$ (defined in~\eqref{eqn:def_l}),
    \begin{align*}
        l_y^{j}(y') :=
        \begin{cases}
            d(h_y, y') & \text{ if } y' < a = q_1 \text{ or } y' > b = q_m \\
            d(y, q_t) & \text{ if } q_t \leq y' < q_{t+1}, 1 \leq t < m  \\
            d(y, q_m) & \text{ if } y' = b = q_m
        \end{cases}
    \end{align*}
    where $h_y \in P_j$, $Q_y = \{q_t\}_t \subset P_j$.
    Hence, for every $y$, $l_y^{j}$ is a \emph{piece-wise linear} function with $O(|Q_y|) = O(\epsilon^{-2})$ (by \cref{claim:m_ub}) pieces,
    where the transition points of $l_y^{j}$ are $Q_y \cup \{0, 1\}$ (noting that $d(h_y, y')$ is linear since $h_y, y' \in P_j$).

    Using that $l$'s are piece-wise linear, we know for $c \in V_i, x \in X \setminus V_i$,
    \begin{align*}
        f_x^{j}(c)
        &= \min_{c', x' \in P_j}\{
            l_c^{j}(c') + d(c', x') + l_x^{j}(x')
        \}  & \text{defined in~\eqref{eqn:def_fxj}}\\
        &= \min_{c', x' \in Q_c \cup Q_x \cup \{0, 1\}}\{
            l_c^{j}(c') + d(c', x') + l_x^{j}(x')
        \}. & \text{as $l$'s are piece-wise linear}
\end{align*}
    Hence, to evaluate $f_x^{j}(c)$ we only need to evaluate
    $l_c^{j}(c')$ and $l_x^j(x')$ at
    $c', x' \in Q_c \cup Q_x \cup \{0, 1\}$,
    and in particular we need to find the piece
    in $l_c^j$ and $l_x^j$ that every $c', x' \in Q_c \cup Q_x \cup \{0, 1\}$
    belong to, and then evaluate a linear function.
    Precisely, the piece that every $c', x'$ belongs to is determined
    by the relative ordering of points $Q_x \cup Q_c$ (recalling that they are from $P_j$).
    Thus, the pieces are not only determined by $x$, but also by $c$
    which is the variable,
    and this means without the information about the pieces,
    $f_x$ cannot be represented as a min-linear function $g_x$.
    Therefore, the idea is to find a partition $\Gamma$ of $V_i$,
    such that for $c$ in each part $V' \in \Gamma$,
    the relative ordering of $Q_c$ with respect to $\{ Q_x : x \in H \}$
    is the same. We note that we need to consider the ordering of $Q_c$
    with respect to all $Q_x$'s, because we care about the relative orderings
    of all $f_x$'s.

    \paragraph{Defining $\Gamma$}
    For $1 \leq j \leq u$, $c \in V_i$, let $\tau_c^{j}$ be the ordering of $Q_c$ with respect to $\bigcup_{y \in H}{Q_y} $ on $P_j$.
    Here, an ordering of $Q_c$ with respect to $\left(\bigcup_{y \in H}{Q_y}\right) $
    is defined by their ordering on $P_j$ which is interpreted as the real line.
    In our definition of $\Gamma$, we will require each part $V' \in \Gamma$
    to satisfy that $\forall c \in V'$,
    the tuple of orderings $(\tau_c^1, \ldots, \tau_c^u )$ remains the same.
    That is, $V_i$ is partitioned according to the joint relative ordering
    $\tau_c^j$'s on all shortest paths $P_j \in \calP_i$.

    Formally, for $1 \leq j \leq u$,
    let $\Lambda^{j} := \{ \tau_c^{j} : c \in V_i \}$ be the collection
    of distinct ordering $\tau_c^j$ on $P_j$ over points $c \in V_i$.
Define
    \begin{align*}
        \Lambda := \Lambda^1 \times \ldots \times \Lambda^u
\end{align*}
    as the tuples of $\tau_j$'s for $ 1\leq j \leq u$ (here, the $\times$ operator is the Cartesian product).
    For $(\tau_1, \ldots, \tau_u) \in \Lambda$,
    define
    \begin{align*}
        V_i^{(\tau_1, \ldots, \tau_u)} := \{ c \in V_i : (\tau_c^1 = \tau_1 )\land \ldots \land (\tau_c^u = \tau_u) \}
    \end{align*}
    as the subset of $V_i$ such that the ordering $\tau_c^j$ for each
    $1\leq j \leq u$ agrees with the given tuple.
    Finally, we define the partition as
    \begin{align*}
        \Gamma := \{ V_i^{(\tau_1, \ldots, \tau_u)} : (\tau_1, \ldots, \tau_u) \in \Lambda \}.
\end{align*}

    \paragraph{Bounding $|\Gamma|$}
By \cref{claim:m_ub}, we know $|Q_y| = O(\epsilon^{-2})$ for every $y \in V$.
    Hence, $\left|\bigcup_{y \in H}{Q_y}\right| = O\left(\epsilon^{-2}|H| \right)$.
    Therefore, for every $j \in [u]$,
    \begin{align*}
        |\Lambda^j|
        \leq \binom{O(\epsilon^{-2}|H|)} { O(\epsilon^{-2}) }
        = O\left(\epsilon^{-1}|H|\right)^{O(\epsilon^{-2})}.
    \end{align*}
    Therefore,
    \begin{align*}
        |\Gamma| \leq \Pi_{1 \leq j \leq u}{ |\Lambda^j| }
        \leq O\left(\epsilon^{-1} |H|\right)^{ O(\epsilon^{-2} u) }
        \leq |H|^{\tilde{O}(\epsilon^{-2}) \cdot u},
    \end{align*}
    as desired.

    \paragraph{Defining $g_x$}
By our definition of $\Gamma$, we need to define
    $g_x$ for each $V' \in \Gamma$.
    Now, fix tuple $(\tau_1, \ldots, \tau_u) \in \Lambda$,
    so the part corresponds to this tuple is
    $V' = V_i^{(\tau_1, \ldots, \tau_u)}$,
    and we will define $g_x$ with respect to such $V'$.
    Similar to the definition of $f_x$'s (see~\eqref{eqn:def_fx}),
    we define $g_x : \mathbb{R}^s \to \mathbb{R}_+$ to have the form
    \begin{align*}
        g_x(y) := \min_{P_j \in \calP_i}{g_x^j(y)}.
    \end{align*}
    Then, for $1 \leq j \leq u$, $x \in H$,
    define
    $g_x^j : \mathbb{R}^s \to \mathbb{R}$ of $s := O(\epsilon^{-2})$ variables
    $(q_1, \ldots, q_m, d(c, q_1), \ldots,$ $d(c, q_m) , h_c)$ for $q_i \in Q_c$, such that
    \begin{align*}
        g_x^j(q_1, \ldots, q_m, d(c, q_1), \ldots, d(c, q_m), h_c )
        = \min_{c', x' \in Q_c \cup Q_x \cup \{0, 1\}}\{
            l_c^j(c') + d(c', x') + l_x^j(x')
            \}.
    \end{align*}
    We argue that for every $1 \leq j \leq u$, $g_x^j$ may be viewed as a minimum of $O(\epsilon^{-4})$ linear functions
    whose variables are the same with that of $g_x^j$.
    \begin{itemize}
        \item Linearity. Suppose $c \in V'$, and fix $c', x' \in Q_c \cup Q_x \cup \{0, 1\}$.
        By the above discussions, $l_c^j(c')$ could take values only from $\{ d(c, q_i) : q_i \in Q_c \} \cup \{ d(h_c, c') \}$.
        Since $\forall q_i \in Q_c$, $d(c, q_i)$ is a variable of $g_x^j$, and $d(h_c, c') = |h_c - c'|$ is linear and that $h_c$ is also a variable of $g_x^j$, we conclude that $l_c^j(c')$
        may be written as a linear function of the same set of variables of $g_x^j$.
        By a similar argument, we have the same conclusion for $l_x^j$. Therefore, $l_c^j(c') + d(c', x') + l_x^j(x')$ may be written as a linear function of $(q_1, \ldots, q_m, d(c, q_1), \ldots, d(c, q_m), h_c)$.
        \item Number of linear functions. By \cref{claim:m_ub}, we have
            \begin{align*}
                \forall y \in V, \qquad |Q_y| = O(\epsilon^{-2}),
            \end{align*}
            hence $|Q_c \cup Q_x \cup \{0, 1\}| = O(\epsilon^{-2})$.
            Therefore, there are $O(\epsilon^{-4})$ pairs of $c', x' \in Q_c \cup Q_x \cup \{0, 1\}$.
    \end{itemize}
    Therefore, item~\ref{itm:gx} of \cref{lemma:min_linear_partition} follows by combining this with the definition of $g_x$.
    We completed the proof of \cref{lemma:min_linear_partition}.
\end{proof}
Now suppose $\Gamma$ is the one that is guaranteed by \cref{lemma:min_linear_partition}.
Since
\begin{align*}
    \left| \{\sigma_c^{\calH} : c \in V_i \} \right|
    \leq \sum_{V' \in \Gamma}{ \left| \{ \sigma_c^{\calH} : c \in V' \} \right| }
\end{align*}
and
\begin{align}
    |\Gamma| \leq |H|^{\tilde{O}(\epsilon^{-2}) \cdot u}
    \leq |H|^{\tilde{O}(\epsilon^{-2})},
    \label{eqn:gamma_ub}
\end{align}
where the last inequality is by \cref{lemma:planar_sep}
(recalling $u = |\calP_i|$),
it suffices to show for every $V' \in \Gamma$,
\begin{align}
    \label{eqn:sigma_goal}
    \left| \{ \sigma_c^{\calH} : c \in V' \} \right| \leq |H|^{\tilde{O}(\epsilon^{-2}) \log{\|X\|_0}}.
\end{align}

Fix some $V' \in \Gamma$.
By \cref{lemma:min_linear_partition}, for every $x \in H$ there exists a min-linear function
$g_x : \mathbb{R}^s \to \mathbb{R}_+$
($s = O(\epsilon^{-2}))$), such that for every $c \in V'$, there exists $y \in \mathbb{R}^s$
that satisfies $f_x(c) = g_x(y)$.
For $y \in \mathbb{R}^s$ define $\pi_y^H$ as a permutation of $H$ that is ordered by
$ g_x(y) $ in non-increasing order and ties are broken in a way that is consistent with $\sigma$.
Then
\begin{align}
    \label{eqn:sigma_pi}
    \left| \{\sigma_c^{\calH_v} : c \in V' \} \right| \leq \left| \{ \pi_y^H : y \in \mathbb{R}^s \} \right|.
\end{align}
We make use of the following lemma to bound the number of permutations $\pi_y^H$.
The lemma relates the number of relative orderings of $g_x$'s to the arrangement
number in computational geometry.
\begin{lemma}[Complexity of Min-linear Functions~\cite{coreset_tw}]
    \label{lemma:min_linear}
    Suppose there are $m$ functions $g_1, \ldots, g_m$ from $\mathbb{R}^{s}$ to $\mathbb{R}$, such that
    $\forall i \in [m]$, $g_i$ is of the form
    \begin{align*}
        g_i(x) := \min_{j \in [t]}\{ g_{ij}(x) \},
    \end{align*}
    where $g_{ij}$ is a linear function.
    For $x \in \mathbb{R}^s$, let $\pi_x$ be the permutation of $[m]$ ordered by
    $g_i(x)$.
    Then,
    \begin{align*}
        \left|\{ \pi_x : x \in \mathbb{R}^s \}\right| \leq (mt)^{O(s)}.
    \end{align*}
\end{lemma}
Applying \cref{lemma:min_linear} on $g_x$'s for $x \in H$ with parameters
$s = O(\epsilon^{-2})$, $t = O(\epsilon^{-4} u) = O\left(\epsilon^{-4}\log{\|X\|_0}\right)$ and $m = |H|$,
we obtain
\begin{align}
    \label{eqn:pi_ub}
    \left| \{ \pi_y^H : y \in \mathbb{R}^s \} \right|
    \leq O\left(\epsilon^{-1} |H| \log{\|X\|_0}\right)^{O(\epsilon^{-2})}
    \leq |H|^{\tilde{O}(\epsilon^{-2}) \cdot \log{\|X\|_0}}.
\end{align}
Thus,~\eqref{eqn:sigma_goal} is implied by combining~\eqref{eqn:pi_ub} with~\eqref{eqn:sigma_pi}.
Finally, we complete the proof of \cref{lemma:planar_sdim}
by combining the above three parts of the arguments.

\subsubsection{From Planar to Minor-excluded Graphs}
\label{sec:planar_to_mf}
The strategy for proving the minor-excluded case is similar to the planar case.
Hence, we focus on presenting the major steps and highlight the differences, while omitting repetitive arguments.
The terminal embedding lemma that we need to prove is restated as follows.
\begin{lemma}[Restatement of Lemma~\ref{lemma:mf_sdim}]
    For every edge-weighted graph $G = (V, E)$ whose shortest path metric is denoted as $M = (V, d)$,
    and every weighted set $X \subseteq V$,
given that $G$ excludes some fixed minor,
    there exists a set of functions $\calF := \{ f_x : V \to \mathbb{R}_+ \mid x \in X \}$ such that
    for every $x \in X$, and $c \in V$, $d(x, c) \leq f_x(c) \leq (1 + \epsilon) \cdot d(x, c)$, and $\sdim_{\max}(\calF) = \tilde{O}(\epsilon^{-2}) \cdot  \log{\|X\|_0}$.
\end{lemma}

Similar to the planar case, we fix $v : X \to \mathbb{R}_+$
and $\calH \subseteq \calF_v$ with $|\calH| \geq 2$ throughout the proof.
Then $\sigma_c^H$ is defined the same as before,
and it suffices to prove
\begin{align*}
    |\{ \sigma_c^\calH : c \in V \}| \leq \poly(\|X\|_0) \cdot |\calH|^{\tilde{O}(\epsilon^{-2}) \log{\|X\|_0}}.
\end{align*}
Next, we used a structural lemma to break $V$ into several parts where each part
is separated by a few shortest paths.
In the planar case, we showed in Lemma~\ref{lemma:planar_sep}
that the number of parts is $O(\|X\|_0)$, and only $O(1)$ separating
shortest paths in $G$ are necessary.
However, the proof of Lemma~\ref{lemma:planar_sep} heavily relies
on planarity, and for minor-excluded graphs, we only manage to prove the following
weaker guarantee.

\begin{lemma}[Structural Property of Minor-excluded Graphs]
    \label{lemma:mf_sep}
    Given edge-weighted graph $G = (V, E)$ that excludes a fixed minor,
    and a subset $S \subseteq V$,
    there is a collection $\Pi := \{ V_i \}_i$ of $V$ with $|\Pi| = \poly(|S|)$ and $\bigcup_{i}{V_i} = V$ such that
    for every $V_i \in \Pi$ the following holds.
    \begin{enumerate}
        \item $|S \cap V_i| = O(1)$.
        \item There exists an integer $t_i$ and $t_i$ groups of paths
        $\calP_1^i, \ldots, \calP_{t_i}^i$ in $G$, such that
        \begin{enumerate}
            \item $|\bigcup_{j=1}^{t_i}{ \calP^i_j }| = O(\log{|S|})$
            \item removing the vertices of all paths in $\bigcup_{j=1}^{t_i}{\calP^i_j}$ disconnects $V_i$ from $V\setminus V_i$ in $G$ (possibly removing points in $V_i$)
            \item for $1 \leq j \leq t_i$, let $G^i_j$ be the sub-graph of $G$
            formed by removing all paths in $\calP^i_1, \ldots, \calP^i_{j-1}$ (define $G^i_1 = G$),
            then every path in $\calP^i_{j}$ is a
            shortest path in $G^i_j$.
        \end{enumerate}
    \end{enumerate}
\end{lemma}
The lemma follows from a recursive application of the balanced shortest path
separator theorem in~\cite[Theorem 1]{DBLP:conf/podc/AbrahamG06}, stated as follows.
\begin{lemma}[Balanced Shortest Path Separator~\cite{DBLP:conf/podc/AbrahamG06}]
    \label{lemma:mf_sp_sep}
    Given edge-weighted graph $G=(V, E)$ that excludes a fixed minor
    with non-negative vertex weight\footnote{\cite[Theorem 1]{DBLP:conf/podc/AbrahamG06} only states the special case with unit vertex weight, while the general weighted version was discussed in a note of the same paper.},
    there is a set of vertices $S \subseteq V$, such that
    \begin{enumerate}
        \item $S = P_1 \cup P_2 \cup \ldots$ where $P_i$ is a set of shortest
        paths in the graph formed by removing $\bigcup_{j < i}{P_j}$
        \item $\sum_{i}{|P_i|} = O(1)$, where the hidden constant depends on the size of the excluded minor
        \item the weight of every component in the graph
        formed by removing $S$ from $G$ is at most half the weight of $V$.
    \end{enumerate}
\end{lemma}
\begin{proof}[Proof of Lemma~\ref{lemma:mf_sep}]
    Without loss of generality, we assume $G$ is a connected graph.
    We will apply Lemma~\ref{lemma:mf_sp_sep} on $G$ recursively
    to define the partition $\Pi$ and
    the groups of shortest paths $\{\calP^i_j\}_j$
    associated with the parts.
    The detailed procedure, called DEF-$\Pi$, is defined in \cref{alg:mf_sep}.
    We assume there is a global $\Gamma$ initialized as $\Gamma = \emptyset$
    which is constructed throughout the execution of the recursive algorithm.
    The execution of the algorithm starts with DEF-$\Pi$($G$, $\emptyset$, $S$).

    Roughly, the procedure DEF-$\Pi$ takes a sub-graph $G'$, a set
    $\mathsf{sep} = \{ \calP_j \}_j$ of groups of paths and $S$ as input,
    such that $G'$ corresponds to a component in a graph formed by removing all paths in $\mathsf{sep}$ from $G$.
    The procedure execute on such $G'$ and find shortest paths
    in $G'$ using Lemma~\ref{lemma:mf_sp_sep}.
    The found shortest paths are segmented (with respect to $S$) and added to the collection $\Pi$.
    Then the found shortest paths are removed from $G'$ to form a new graph $G''$.
    Components in $G''$ that contain less than $2$ points in $S$ are made
    new parts in $\Pi$, and the procedure DEF-$\Pi$ is invoked recursively
    on other components in $G''$.

    \begin{algorithm}[ht]
        \caption{Procedure DEF-$\Pi$($G' = (V', E')$, $\mathsf{sep}$, $S$)}
        \label{alg:mf_sep}
        \begin{algorithmic}[1]
            \State apply Lemma~\ref{lemma:mf_sp_sep} on graph $G'$
            with vertex weight $1$ if $x \in V' \cap S$
            and $0$ otherwise, and let $\calP$ be
            the set of shortest paths in $G'$ guaranteed by the lemma.
            \For{$P \in \calP$}
                \State interpret $P$ as interval $[0, 1]$,
                list $S \cap P =\{x_1, \ldots, x_m\}$
                and $0 \leq x_1 \leq \ldots \leq x_m \leq 1$
                \State segment $P$ into sub-paths
                $\calP' = \{[0, x_1],
                [x_1, x_2], \ldots, [x_m, 1] \}$
                \For{$P' \in \calP'$}
                    \State include $P'$ in $\Pi$, and define the set
                    of associated groups of shortest paths as $\mathsf{sep} \cup \{ P' \}$
                \EndFor
            \EndFor
            \State let $G''$ be the graph formed by removing all paths in $\calP$, and let $\mathcal{C} = \{ C_i \}_i$ be its components
            \State include the union of all components with no intersection with $S$ as a single part in $\Pi$, and define
            the set of associated groups of paths as $\mathsf{sep} \cup \calP$
            \For{$C_i \in \mathcal{C}$}
                \If{$|C_i \cap S| = 1$}
                    \State include $C_i$ as a new part in $\Pi$, and define the set of associated groups of paths as $\mathsf{sep} \cup \calP$
                \ElsIf{$|C_i \cap S| \geq 2$}
                    \State call DEF-$\Pi$($G''[C_i]$, $\mathsf{sep} \cup \{ \calP \}$, $S$) \Comment{$G''[C_i]$ is the induced sub-graph of $G''$ on vertex set $C_i$}
                \EndIf
            \EndFor
        \end{algorithmic}
    \end{algorithm}

    By construction and Lemma~\ref{lemma:mf_sp_sep},
    it is immediate that $\bigcup_{V_i \in \Pi}{V_i} = V$,
    and item 2.(b), 2.(c) also follows easily.
    To see item 1, we observe that we have two types of $V_i$'s in $\Pi$.
    One is from the shortest paths $\calP$ (Line 6), and because of the segmentation,
    the intersection with $S$ is at most $2$.
    The other type is the components in $G''$
    whose intersection with $S$ is by definition at most $1$ (Line 10, 13).
    Therefore, it remains to upper bound $|\Pi|$, and show item 2.(a) which
    requires a bound of $|\bigcup_{j=1}^{t_i}{\calP^i_j}| = O(\log{|S|})$ for all $V_i \in \Pi$.

    First, we observe that at any execution of Gen-$\Pi$,
    it is always the case that $0 \leq |\mathsf{sep}| \leq O(\log{|S|})$,
    because Lemma~\ref{lemma:mf_sp_sep} guarantees the weight of every component in $G''$ is halved.
    This also implies that the total number of executions of GEN-$\Pi$ is $\poly(|S|)$.
    Therefore, $\forall V_i \in \Pi$, $|\bigcup_{j=1}^{t_i}{\calP^i_j}|
    \leq O(\log{|S|})$,
    which proves item 2.(a).

    \paragraph{Bounding $|\Pi|$}
    Observe that there are three places where we include a part $V_i$ in $\Pi$,
    and we let $\Pi_1$ be the subset of those included at Line 6,
    $\Pi_2$ be those included at Line 10, and $\Pi_3$ be those included at Line 13.
    Then $|\Pi| \leq |\Pi_1| + |\Pi_2| + |\Pi_3|$.

    If $V_i \in \Pi_1$, then $V_i$ is a sub-path of some $P \in \calP$,
    where $\calP$ is defined at Line 1.
    We observe that the number of all $V_i \in \Pi_1$ such that
    $V_i \cap S \neq \emptyset$,
    i.e. $|\{ V_i \in \Pi_1 : V_i \cap S \neq \emptyset \}|$,
    is at most $O(|S|)$. This is because we remove paths $P \in \calP$
    in every recursion,
    which means any point in $S$ can only participate in at most one such $P$
    during the whole execution,
    and hence any point in $S$ can intersect at most two sub-paths $V_i \in \Pi_1$
    such that $V_i \cap S \neq \emptyset$ (because $|V_i \cap S| \leq 2$
    by the segmentation at Line 4).
    On the other hand, if $V_i \in \Pi_1$ and $V_i \cap S = \emptyset$,
    then no segmentation was performed and $V_i = P$ for $P$ at Line 2.
    Therefore, the number of such $V_i$'s is bounded by the total number of
    execution of DEF-$\Pi$ multiplied by the size of $\calP$ at Line 2,
    which is at most $\poly(|S|)$.
    Therefore, we conclude that $|\Pi_1| = \poly(|S|)$.

    Finally,
    since every $V_i \in \Pi_3$ satisfies $|V_i \cap S| = 1$ (at Line 12 and 13),
    and we observe that subsets in $\Pi_3$ are disjoint,
    so we immediately have $|\Pi_3| = O(|S|)$.
    For $\Pi_2$, we note that only one $V_i \in \Pi_2$ could be included
    in each execution of DEF-$\Pi$, so $|\Pi_2| = \poly(|S|)$.

    We conclude the proof of Lemma~\ref{lemma:mf_sep} by combining all the above discussions.
\end{proof}
As before, we still apply the Lemma~\ref{lemma:mf_sep} with $S = X$ (which is unweighted set) to
obtain $\Gamma = \{V_i\}_i$ with $|\Pi| = O(\poly(\|X\|_0))$,
and it suffices to prove for each $V_i \in \Pi$
\begin{align*}
    |\{ \sigma_c^\calH : c \in V_i \}| \leq |\calH|^{\tilde{O}(\epsilon^{-2}) \log{\|X\|_0}}.
\end{align*}
To proceed, we fix $V_i$ and define functions $f_x(\cdot)$ for $c \in V_i$.
However, compared with Lemma~\ref{lemma:planar_sep},
the separating shortest paths in Lemma~\ref{lemma:mf_sep} are not from the original graph $G$,
but is inside some sub-graph generated by removing various other
separating shortest paths.
Also, the number of shortest paths in the separator
is increased from $O(1)$ to $O(\log{\|X\|_0})$.

Hence, we need to define $f_x$'s with respect to the new structure of the separating shortest paths.
Suppose $\{ \calP^i_1, \ldots, \calP^i_{t_i} \}$ is the $t_i$ groups
of paths guaranteed by Lemma~\ref{lemma:mf_sep}. Also as in the lemma,
suppose $G^i_j$ is the sub-graph of $G$ formed by removing all paths in
$\calP^i_1, \ldots, \calP^i_{j-1}$ (define $G^i_1 = G$).
For $1 \leq j \leq t_i$, $P \in \calP^i_j$ and $x, y \in V$,
let $d_j^P(x, y)$ denote the length of the shortest path from $x$ to $y$
using edges in $G^i_j$ and uses at least one point of $P$.
Then, analogue to Fact~\ref{fact:sp_cross}, we have the following lemma.
\begin{lemma}
    \label{lemma:mf_sp_cross}
    For $c \in V_i$ and $x \in V \setminus V_i$,
    there exists $1 \leq j \leq t_i$, $P \in \calP^i_j$ and $c', x' \in P$,
    such that $d(c, x) = d^P_j(c, c') + d^P_j(c', x') + d^P_j(x', x)$.
\end{lemma}
\begin{proof}
    First, we observe that the shortest path $c \rightsquigarrow x$
    has to intersect (at a vertex of) at least one path contained in $\{\calP^i_j\}_j$,
    because removing $\bigcup_{j=1}^{t_i}{\calP_j}$ disconnects $V_i$
    from $V \setminus V_i$.
    Suppose $j_0$ is the smallest $j$ such that $c \rightsquigarrow x$
    intersects a shortest path in $\calP^i_j$, and let $P \in \calP^i_{j_0}$
    be any intersected path in $\calP^i_{j_0}$.

    Then, this implies that (the edge set of) $c \rightsquigarrow x$
    is totally contained in sub-graph $G^i_{j_0}$,
    since $G^i_{j_0}$ is formed by removing only groups $\calP^s_j$
    with $j < j_0$ which do not intersect $c \rightsquigarrow x$.
    Hence, we have $d(c, x) = d_{G^i_{j_0}}(c, x)$, where
    $d_{G^i_{j_0}}$ is the shortest path metric in sub-graph $G^i_{j_0}$.
    By Lemma~\ref{lemma:mf_sep}, $P$ is a shortest path in $G^i_{j_0}$,
    so $c \rightsquigarrow x$ has to cross $P$ at most once, which implies
    there exists $c', x' \in P$, such that $d(x, c) = d^P_j(c, c')
    + d^P_j(c', x') + d^P_j(x', x)$, as desired.
\end{proof}
Using Lemma~\ref{lemma:mf_sp_cross} and by the optimality of the shortest path,
we conclude that
\begin{align*}
    \forall c \in V_i, x \in X, \quad
    d(c, x) = \min_{1 \leq j \leq t_i}{\min_{P \in \calP^i_j}{\min_{c', x' \in P}
    \{ d^P_j(c, c') + d^P_j(c', x') + d^P_j(x', x) \}
    } }.
\end{align*}
Then, for each $1 \leq j \leq t_i$, path $P \in \calP^i_j$,
we use the same way as in the planar case to define
the approximate distance function $l$
to approximate $d^P_j(y, y')$ for $y \in V$ and $y' \in P$.
The $f_x$ is then defined similarly, and the distortion follows
by a very similar argument as in Lemma~\ref{lemma:l_error}.

The analysis of shattering dimension is also largely the same
as before, except that the definition of $u$ in the statement of
Lemma~\ref{lemma:min_linear_partition} is slightly changed because of the new structural lemma.
The new statement is presented as follows, and the proof of it is essentially as before.
\begin{lemma}
    Let $u = |\bigcup_{j = 1}^{t_i}{\calP^i_j}|$.
    There exists a partition $\Gamma$ of $V_i$, such that the following holds.
    \begin{enumerate}
        \item $|\Gamma| \leq |H|^{\tilde{O}(\epsilon^{-2})\cdot u}$.
        \item $\forall V' \in \Gamma$, $\forall x \in H$,
        there exists $g_x : \mathbb{R}^s \to \mathbb{R}_+$ where $s = O(\epsilon^{-2})$,
        such that $g_x$ is a minimum of $O(\epsilon^{-4} u)$ linear functions on $\mathbb{R}^s$,
        and for every $c \in V'$, there exists $y \in \mathbb{R}^s$ that satisfies $f_x(c) = g_x(y)$.
    \end{enumerate}
\end{lemma}
We apply the lemma with the new bound of
$u = |\bigcup_{j = 1}^{t_i}{\calP^i_j}| = O(\log{\|X\|_0})$ (by Lemma~\ref{lemma:mf_sep}),
and the bound in~\eqref{eqn:sigma_goal} is increased to
\begin{align*}
    |\Gamma| \leq |H|^{\tilde{O}(\epsilon^{-2}) \cdot u}
    \leq |H|^{\tilde{O}(\epsilon^{-2}) \log{\|X\|_0}}.
\end{align*}
Finally, to complete the proof of Lemma~\ref{lemma:mf_sdim},
we again use Lemma~\ref{lemma:min_linear} on each $V' \in \Gamma$
to conclude the desired shattering dimension bound.

\subsection{High-Dimensional Euclidean Spaces}
\label{sec:Euclidean}

We present a terminal embedding for Euclidean spaces,
with a guarantee that is similar to that of excluded-minor graphs. 
For these results, the ambient metric space $(V,d)$ of all possible centers
is replaced by a Euclidean space.\footnote{It is easily verified that as long as $X$ is finite,
  our entire framework from \cref{sec:framework} 
  extends to $V=\RR^m$ with $\ell_2$ norm. 
  For example, all maximums (e.g., in \cref{lemma:generalized_fl})
  are well-defined by using compactness arguments on a bounding box. 
}

\begin{lemma}
\label{lemma:euclidean_em}
For every $\epsilon \in (0, 1/2)$ and finite weighted set $X \subset \RR^m$, 
there exists $\calF = \{ f_x : \RR^m \to \RR_+ \mid x \in X \}$ 
such that
\[
  \forall x \in X, c \in \mathbb{R}^m,
  \qquad
  \|x - c\|_2 \leq f_x(c) \leq (1 + \epsilon) \|x - c\|_2,
\]
and $\sdim_{\max}(\calF) = O(\epsilon^{-2}\log{\| X \|_0})$.
\end{lemma}

\begin{proof}
The lemma follows immediately from the following 
terminal version of the \JL Lemma~\cite{JL84}, 
proved recently by Narayanan and Nelson~\cite{DBLP:conf/stoc/NarayananN19}. 

\begin{theorem}[Terminal \JL Lemma~\cite{DBLP:conf/stoc/NarayananN19}]
\label{thm:terminal_jl}
For every $\epsilon \in (0, 1/2)$ and finite $S \subset \mathbb{R}^m$,
there is an embedding $g : S \to \mathbb{R}^t$
for $t = O(\epsilon^{-2}\log{|S|})$,
such that
\begin{align*}
  \forall x \in S, y \in \mathbb{R}^m,\quad
  \| x - y \|_2 \leq \| g(x) - g(y) \|_2 \leq (1+\epsilon) \| x - y \|_2.
\end{align*}
\end{theorem}
  
Given $X\subset \RR^m$,
apply \cref{thm:terminal_jl} with $S = X$ (as an unweighted set),
and define for every $x \in X$ the function $f_x( c ) := \| g(x) - g(c) \|_2$.
Then $\calF = \{ f_x \mid x \in X \}$ clearly satisfies the distortion bound.
The dimension bound follows by plugging $t = O(\epsilon^{-2}\log{\|X\|_0})$
into the bound $\sdim_{\max} (\calF) = O(t)$
known from~\cite[Lemma 16.3]{DBLP:conf/stoc/FeldmanL11}.\footnote{The following is proved
  in~\cite[Lemma 16.3]{DBLP:conf/stoc/FeldmanL11}. 
For every $S \subset \RR^t$,
the function set $\calH := \{ h_x \mid x \in S \}$
given by $h_x(y) = \|x - y\|_2$,
has shattering dimension $\sdim_{\max}(\calH) = O(t)$. 
}
\end{proof}

\begin{corollary}[Coresets for Euclidean Spaces]
\label{cor:coreset_euclidean}
For every $0 < \epsilon, \delta < 1/2$, $z \geq 1$, and integers $k, m \geq 1$,
Euclidean \kzC of every weighted set $X \subset \mathbb{R}^m$
admits an $\epsilon$-coreset of size
$\tilde{O}(\epsilon^{-4} 2^{2z} k^2 \log{\frac{1}{\delta}}) $.
Furthermore, such a coreset can be computed\footnote{We assume that evaluating $\|x - y\|_2$ for $x, y \in \mathbb{R}^m$
  takes time $O(m)$.
}
in time $\tilde{O}(k \|X\|_0 m)$ with success probability $1 - \delta$.
\end{corollary}

\begin{proof}
By combining \cref{lemma:framework_mult}, \cref{lemma:alg_2_time}
with our terminal embedding from \cref{lemma:euclidean_em},
we obtain an efficient algorithm for constructing a coreset
of size $\tilde{O}(\epsilon^{-4} 2^{2z} k^2 \log{\|X\|_0})$. 
This size can be reduced to the claimed size (and running time) 
using the iterative size reduction of \cref{thm:ite_size_reduct}.
\end{proof}

\begin{remark}[Comparison to~\cite{HV20}]
\label{remark:euclidean_coreset}
For \kzC in Euclidean spaces, our algorithms can also compute
an $\epsilon$-coreset of size $\tilde{O}(\epsilon^{-O(z)}k)$,
which offers a different parameters tradeoff than \cref{cor:coreset_euclidean}. 
This alternative bound is obtained by simply replacing the application of
\cref{lemma:generalized_fl} (which is actually from~\cite{fss13})
with~\cite[Lemma 3.1]{HV20} (which itself is a result from~\cite{DBLP:conf/stoc/FeldmanL11}, extended to weighted inputs). 

Our two coreset size bounds are identical to
the state-of-the-art bounds proved by Huang and Vishnoi~\cite{HV20} (in the asymptotic sense).
Their analysis is different, and bounds $\sdim_{\max}$ independently of $X$
using a dimensionality-reduction argument for clustering objectives. 
In contrast, we require only a loose bound
$\sdim_{\max}(\calF) = O(\poly(\epsilon^{-1})\cdot \log{\|X\|_0})$,
which follows immediately from~\cite{DBLP:conf/stoc/NarayananN19},
and the coreset size is then reduced iteratively
using \cref{thm:ite_size_reduct},
which simplifies the analysis greatly.
\end{remark}

\subsection{Graphs with Bounded Highway Dimension}
\label{sec:hw}

The notion of highway dimension was proposed by
Abraham, Fiat, Goldberg, and Werneck~\cite{DBLP:conf/soda/AbrahamFGW10}
to measure the complexity of road networks. 
Motivated by the empirical observation that a shortest path between 
two far-away cities always passes through a small number of hub cities,
the highway dimension is defined, roughly speaking,
as the maximum size of a hub set that meets every long shortest path, 
where the maximum is over all localities of all distance scale. 
Several slightly different definitions of highway dimension appear in the literature, and we use the one proposed in~\cite{FFKP18}.

\begin{definition}[Highway Dimension~\cite{FFKP18}]
    \label{def:hdim}
    Fix some universal constant $\rho \geq 4$.
    The highway dimension of an edge-weighted graph $G=(V, E)$, denoted $\hdim(G)$,
    is the smallest integer $t$ such that for every $r \geq 0$ and $x \in V$,
    there is a subset $S \subseteq B(x, \rho r)$ with $|S| \leq t$,
    such that $S$ intersects every shortest path of length at least $r$
    all of whose vertices lie in $B(x, \rho r)$.
\end{definition}
\begin{remark}
  \label{remark:highway}
  This version generalizes the original one from~\cite{DBLP:conf/soda/AbrahamFGW10}
  (and also the subsequent journal version~\cite{DBLP:journals/jacm/AbrahamDFGW16}),
  and it was shown to capture a broader range of real-world transportation networks~\cite{FFKP18}.
  We also note that the version in~\cite{DBLP:journals/jacm/AbrahamDFGW16}
  is stronger than the notion of doubling dimension~\cite{DBLP:conf/focs/GuptaKL03}, however, the version that we use (from~\cite{FFKP18}) is not.
  In particular, it means that the previous coreset result for doubling metrics~\cite{DBLP:conf/focs/HuangJLW18} does not apply to our case.
\end{remark}

Unlike the excluded-minor and Euclidean cases mentioned in earlier sections,
our coresets for graphs with bounded highway dimension are obtained
using terminal embeddings with an additive distortion.

\begin{lemma}
\label{lemma:hw_sdim}
Let $G = (V, E)$ be an edge-weighted graph and denote its shortest-path metric by $M(V, d)$.
Then for every $0 < \epsilon < 1/2$, weighted set $X \subseteq V$
and an (unweighted) subset $S \subseteq V$,
there exists $\calF_S = \{ f_x : V \to \mathbb{R}_+ \mid x \in X \}$
such that
\begin{align*}
  \forall x \in X, c \in V, \quad
  d(x, c) \leq f_x(c)  \leq (1+\epsilon) \cdot d(x, c) + \epsilon\cdot
  d(x, S),
\end{align*}
and
$\sdim_{\max}(\calF_S) = \left(|S| + \hdim(G)\right)^{O(\log(1/\epsilon))}$. 
\end{lemma}

\begin{proof}
We rely on an embedding of graphs with bounded highway dimension
into graphs with bounded treewidth, as follows.
\begin{lemma}[\cite{DBLP:conf/esa/BeckerKS18}]
\label{lemma:highway_tw}
For every $0 < \epsilon < 1/2$,
edge-weighted graph $G=(V, E)$ of highway dimension $h$, and $S \subseteq V$,
there exists a graph $G'= (V', E')$
of treewidth $\tw(G') = (|S| + h)^{O(\log(1/\epsilon))}$,
and a mapping $\phi : V \to V'$ such that 
\[
  \forall x, y \in V, \quad
  d_G(x, y)
  \leq d_{G'}(\phi(x), \phi(y))
  \leq (1+\epsilon) \cdot d_G(x, y) + \epsilon \cdot \min\{ d(x, S), d(y, S) \}.
\]
\end{lemma}

We now apply on $G'$ (the graph produced by \cref{lemma:highway_tw}),
the following result from~\cite[Lemma 3.5]{coreset_tw},
which produces the function set $\calF_S$ we need for our proof.

\begin{lemma}[\cite{coreset_tw}]
\label{lemma:sdim_tw}
Let $G = (V, E)$ be an edge-weighted graph,
and denote its shortest-path metric by $M(V, d)$.
Then for every weighted set $X \subseteq V$,
the function set $\calF = \{ d(x, \cdot) \mid x \in X \}$
has $\sdim_{\max}(\calF) = O(\tw(G))$,
where $\tw(G)$ is the treewidth of $G$.
\end{lemma}

Notice that we could also apply on $G'$ our own \cref{lemma:mf_sdim}, 
because bounded-treewidth graphs are also excluded-minor graphs,
however \cref{lemma:sdim_tw} has better dependence on $\tw(G)$
and also saves a $\poly(1/\epsilon)$ factor.
This concludes the proof of \cref{lemma:hw_sdim}. 
\end{proof}

\begin{corollary}[Coresets for Graphs with Bounded Highway Dimension] 
    \label{cor:coreset_hw}
    For every edge-weighted graph $G = (V, E)$, $0 < \epsilon, \delta < 1/2$,
    and integer $k \geq 1$, \kMedian of every weighted set $X \subseteq V$
    (with respect to the shortest path metric of $G$) admits
    an $\epsilon$-coreset of size
    $\tilde{O}((k + \hdim(G))^{O(\log(1/\epsilon))}) \log{\frac{1}{\delta}})$.
    Furthermore, it can be computed in time $\tilde{O}(|E|)$
    with success probability $1 - \delta$.
\end{corollary}

\begin{proof}
By combining \cref{lemma:framework_add}, \cref{cor:alg_3_time}
with our terminal embedding from \cref{lemma:hw_sdim}, 
we obtain an efficient also for constructing a coreset of the said size. 
Notice that we do not need to apply the iterative size reduction (\cref{thm:ite_size_reduct})
because $\sdim_{\max}$ is independent of $X$,
thanks to the additive error.
\end{proof}

 \section{Applications: Improved Approximation Schemes for \kMedian}
\label{sec:application}
In this section, we apply coresets to design approximation schemes for \kMedian in
shortest-path metrics of planar graphs and graphs with bounded highway dimension.
In particular, we give an FPT-PTAS, parameterized by $k$ and $\epsilon$,
for \kMedian in graphs with bounded highway dimension,
and a PTAS for \kMedian in planar graphs.
Both algorithms run in time near-linear in $|V|$ and improve state of the art results.

\paragraph{FPT-PTAS}
An $\epsilon$-coreset $D$ reduces the size of the input data set $X$ while approximately preserving the cost for all clustering centers.
Intuitively, in order to find a $(1 + \epsilon)$-approximate solution, it suffices to solve the problem on $D$ instead of $X$.
However, solving the problem on $D$ does not necessarily imply a PTAS for $X$
because the optimal center $C$ maybe contain element from the ambient space $V$,
and thus would require enumerating all center sets from $V^k$ making this approach prohibitively expensive.
Instead, we enumerate all $k$-partitions of $D$ and find an optimal center for each part.
This simple idea implies an FPT-PTAS for \kMedian and it can be implemented efficiently if
the coreset size is independent of the input $X$. We formalize this idea in \cref{sec:fpt_ptas}.

\paragraph{Centroid Set}
The aforementioned simple idea of enumerating all $k$-partitions of the coreset has exponential
dependence in $k$, and hence is not useful for PTAS.
Precisely, the bottleneck is that the set of potential centers, which is $V$, is not reduced.
To reduce the potential center set, we consider \emph{centroid set} that was first introduced by~\cite{DBLP:journals/dcg/Matousek00} in the Euclidean setting, and later has been extended to other settings, e.g., doubling spaces~\cite{DBLP:conf/focs/HuangJLW18}.
A centroid set is a subset of $V$ that contains a $(1 + \epsilon)$-approximate solution.
We obtain centroid sets of size \emph{independent} of the input $X$ for planar \kMedian, improving the recent bound of $(\log{|V|})^{\epsilon^{-O(1)}}$ from~\cite{DBLP:conf/esa/Cohen-AddadPP19}.
The formal statement of our result for the centroid set can be found in \cref{sec:centroid}.

\paragraph{PTAS for Planar \kMedian}
The aforementioned improvement for centroid sets immediately implies improved PTAS for \kMedian.
Indeed, a $(1 + \epsilon)$-approximation for the centroid set is as well
a $(1 + \Theta(\epsilon))$-approximation for the original data set.
Specifically, we apply our centroid set to speedup a local search algorithm~\cite{cohen2019local} for planar \kMedian,
and our result is a PTAS that runs in time $\tilde{O}\left((k\epsilon^{-1})^{\epsilon^{-O(1)}} |V|\right)$ which is near-linear in $|V|$.
This improves a previous PTAS~\cite{cohen2019local} whose running time
is $k^{O(1)} |V|^{\epsilon^{-O(1)}}$,
and an FPT-PTAS~\cite{DBLP:conf/esa/Cohen-AddadPP19} whose running time
is $2^{O(k\epsilon^{-3}\log(k\epsilon^{-1}))} |V|^{O(1)}$.
Details of the PTAS can be found in \cref{sec:ptas_planar}.

\subsection{FPT-PTAS}
\label{sec:fpt_ptas}
We state our FPT-PTAS as a general reduction. Specifically we show that if
a graph family admits a small $\epsilon$-coreset then it also admits an efficient FPT-PTAS.
\begin{lemma}
    \label{lemma:fpt_ptas}
    Let $\calG$ be family of graphs.
    Suppose for $0 < \epsilon < \frac{1}{2}$, integer $k \geq 1$, every graph $G = (V, E) \in \calG$
    and every weighted set $X \subseteq V$, there is an $\epsilon$-coreset
    $D = D(G, X)$ for \kMedian on $X$ in the shortest-path metric of $G$.
    Then there exists an algorithm
    that for every $0 < \epsilon < \frac{1}{2}$, integer $k \geq 1$ and $G \in \calG$
    computes a $(1+\epsilon)$-approximate solution for \kMedian
    on any weighted set $X \subseteq V$ in time $\tilde{O}(k^{1 + \| D(G, X) \|_0}  |V|)$.
\end{lemma}
\begin{proof}
    The algorithm finds an \emph{optimal} solution for
    the weighted instance defined by $D$. This optimal solution is a $(1 + \epsilon)$-approximate
    solution for \kMedian on the original data set $X$ since $D$ is an $\epsilon$-coreset.

    To find the optimal solution for \kMedian on $D$ we enumerate all $k$-clusterings (i.e. $k$-partitions)
    $\mathcal{C} = \{ C_1, \ldots, C_k \}$ of $D$. For each part $C_i$ we find an optimal
    center $c_i \in V$ that minimizes the cost of part $C_i$,
    i.e. $\min_{c_i \in V}{ \sum_{x \in C_i}{w_D(x) \cdot d(x, c_i)} }$.
    The optimal solution is the $k$-center set that achieves the minimum
    total cost over all such $k$-clusterings of $D$.

    To implement this algorithm efficiently,
    we first pre-compute all distances between point in $D$ and points in $V$. This can be done in time $\tilde{O}(\|D\|_0 |V|)$ using e.g. Dijkstra' algorithm. Using the pre-computed distances, we can find, in $O(|V|)$ time,
    the optimal center for any fixed set $C \subseteq D$. Since there are $k$ parts $C_1,\dots, C_k$ and since there are $k^{\|D\|_0}$ possible partitions, the total running time is $\tilde{O}\left(k^{1 + \|D\|_0} |V|\right)$. This completes the proof.
\end{proof}

\paragraph{FPT-PTAS for Graphs with Bounded Highway Dimension}
Combining \cref{lemma:fpt_ptas} with \cref{cor:coreset_hw},
we obtain an FPT-PTAS for \kMedian in graphs of bounded highway dimension.
Compared with the previous bound $|V|^{O(1)} \cdot f(\epsilon, k, \hdim(G))$ from~\cite[Theorem 2]{DBLP:conf/esa/BeckerKS18},
our result runs in time near-linear in $|V|$ which is a significant improvement.
Moreover, our algorithm is based on straightforward enumeration while~\cite{DBLP:conf/esa/BeckerKS18} is based on dynamic programming.
\begin{corollary}
    \label{cor:fpt_ptas_hw}
    There is an algorithm that for every $0 < \epsilon < \frac{1}{2}$,
    integer $k \geq 1$, every edge-weighted graph $G=(V, E)$,
    computes a $(1 + \epsilon)$-approximate solution for \kMedian
    on every weighted set $X \subseteq V$ with constant probability,
    running in time $\tilde{O}\left(|V| \cdot k^{(k + \hdim(G))^{O(\log{\epsilon^{-1}})}}\right)$.
\end{corollary}
Similarly, plugging \cref{cor:coreset_mf} into \cref{lemma:fpt_ptas}
yields an FPT-PTAS for \kMedian in planar graphs.
We do not state this result here because the improved PTAS in the following section has a better running time.

\subsection{Centroid Sets}
\label{sec:centroid}
The focus of the section is to present an improved centroid set that
will be combined with a local search algorithm to yield a better PTAS.
As already mentioned, a centroid set is a subset of points
that contains a near-optimal solution. The formal definition is given below,
and our centroid set is presented in \cref{thm:centroid_planar}.
\begin{definition}[Centroid Set]
    \label{def:centroid_set}
    Given a metric space $M(V, d)$ and weighted set $X \subseteq V$,
    a set of points $S \subseteq V$ is an $\epsilon$-centroid set
    for \kzC on $X$ if there is a center set $C \in S^k$ such that
    $\cost_z(X, C) \leq (1 + \epsilon) \cdot \OPT_z(X)$.
\end{definition}

\begin{theorem}
    \label{thm:centroid_planar}
    There is an algorithm that
    computes an $\epsilon$-centroid set $D$ of size
    \begin{align*}
        \|D\|_0 = (\epsilon^{-1})^{O(\epsilon^{-2})} \poly(\| X \|_0),
    \end{align*}
    for every $0 < \epsilon < \frac{1}{2}$, every planar
    graph $G = (V, E)$ and weighted subset $X \subseteq V$,
    running in time $\tilde{O}((\epsilon^{-1})^{O(\epsilon^{-2})}\poly(\|X\|_0)|V|)$.
\end{theorem}

First of all, we show there is a near-optimal
solution $C^\star$ such that the distance from every center in $C^\star$
to $X$ can only belong to $\poly(\|X\|_0)$ number of distinct distance scales.
This is an essential property to achieve centroid sets of size
independent of $V$.
Specifically, consider the pairwise distance between points in $X$,
and assume they are sorted as
$$
d_1\leq d_2\leq \ldots \leq d_m
$$
where $m=\binom{\|X\|_0}{2}$. We prove the following lemma.

\begin{lemma} \label{dislabel}
For every $\epsilon\in (0,1/2)$, there is a $k$-subset $C\subseteq V^k$ and an assignment
$\pi: X \to C$, such that
\begin{align}
    \sum_{x\in X} {w_X(x) \cdot d(x,\pi(x))}
    \leq (1+2\epsilon) \cdot \OPT(X), \label{eqn:dist_pair}
\end{align}
and for every $x\in X$, $d(x,\pi(x))$ belongs to an interval
$\mathcal{I}_j:=[\epsilon d_j,d_j/\epsilon]$ for some $j=1, \ldots, m$. In particular, $C$ is a $(1+2\epsilon)$-approximation to \kMedian on $X$.
\end{lemma}

\begin{proof}
Let $C^\star = \{c_1^\star, \ldots , c_k^\star \} \subseteq V$ be the
optimal solution to \kMedian on $X$, and we will define $C$ by ``modifying'' $C^\star$.
Let $C_i^\star \subseteq X$ be the corresponding cluster of $c_i^\star$ and
define $\cost_i^\star :=\sum_{x\in C_i^\star} {w_X(x) \cdot d(x, c_i^\star)}$ to be the cost contributed by $C^\star_i$.

The proof strategy goes as follows.
We examine $c^\star_i \in C^\star$ one by one.
For each $c^\star_i$, we will define $c_i \in C$ as some point in $C^\star_i$,
and the assignment $\pi$ assigns every point in $C^\star_i$ to $c_i$.
To bound the cost, we will prove
$\sum_{x \in C^\star_i}{w_X(x) \cdot d(x, c_i)} \leq (1 + 2\epsilon) \cdot \cost^\star_i$ for each $i$,
and this implies~\eqref{eqn:dist_pair}.

Now fix some $i$.
If $c^\star_i$ satisfies for every $x \in X$,
there is some $1 \leq j \leq m$ such that $d(x, c^\star_i)$
belongs to $\mathcal{I}_j = [\epsilon d_j, \epsilon^{-1} d_j]$,
then we include $c_i := c^\star_i$ to $C$,
and for all $x \in C^\star_i$, let $\pi(x) := c^\star_i$.
Since the center $c^\star_i$ is included in $C$ as is,
the cost corresponding to $C^\star_i$ is not changed.

Otherwise, there is some $\hat{x} \in X$ such that
for every $1 \leq j \leq m$, either $d(\hat{x}, c^\star_i) < \epsilon d_j$
or $d(\hat{x}, c^\star_i) > \epsilon^{-1}d_j$.
Then we pick any such $\hat{x}$, let $c_i:=\hat{x}$,
and define for each $x \in C^\star_i$, $\pi(x) := \hat{x}$.
We note that for every $x'\in X$, $d(\hat{x},x')$ equals some $d_j$ by definition, so $d(\hat{x},x')=d_j\in \mathcal{I}_j$.

Hence, it remains to prove that the cost is still bounded, i.e.
$\sum_{x \in C^\star_i}{w_X(x) \cdot d(x, \hat{x})} \leq (1 + 2\epsilon) \cdot \cost_i^\star$,
and we prove it by showing $\forall x \in C^\star$, $d(x, \hat{x}) \leq (1 + 2\epsilon) \cdot d(x, c^\star_i)$.
Observe that $d(x, \hat{x}) = d_j$ for some $j$,
so depending on whether $d(\hat{x}, c^\star_i) < \epsilon d_j$ or $d(\hat{x}, c^\star_i) > \epsilon^{-1} d_j$ we have two cases.
\begin{itemize}
    \item If $d(\hat{x} , c^\star_i) < \epsilon d_j=\epsilon d(x, \hat{x})$,
    then by triangle inequality, $d(x, \hat{x}) \leq d(x, c^\star_i) + d(c^\star_i, \hat{x})\leq d(x, c^\star_i)+\epsilon d(x, \hat{x})$, hence $d(x,\hat{x})\leq \frac{1}{1-\epsilon} d(x,c^\star_i)\leq (1+2\epsilon) d(x,c^\star_i)$ when $\epsilon\in (0,1/2)$.
    \item Otherwise, $d(\hat{x}, c^\star_i) > \epsilon^{-1} d_j$, by triangle inequality,
    $d(x, c^\star_i)\geq d(\hat{x}, c^\star_i)-d(x, \hat{x})>\frac{1-\epsilon}{\epsilon} d(x, \hat{x})$,
    which implies $d(x,\hat{x})<\frac{\epsilon}{1-\epsilon}d(x, c^\star_i)<(1+\epsilon) d(x,c^\star_i)$ when $\epsilon\in (0,1/2)$.
\end{itemize}
This completes the proof.
\end{proof}

\begin{proof}[Proof of \cref{thm:centroid_planar}]
Suppose $C^\star$ is an optimal solution.
Our general proof strategy is to find a point $c'$ that is sufficiently close to $c$ for very center point $c \in C^\star$.
Specifically, consider a center point $c \in C^\star$, and let $x_c \in X$
be the \emph{closest} point to it.
We want to guarantee that there always exists some $c'$ in the centroid set,
such that $d(c,c')\leq \epsilon \cdot d(c,x_c)$,
and this would imply the error guarantee of the centroid set by triangle inequality.

Since we can afford $(1 + \epsilon)$-multiplicative error,
we round the distances to the nearest power of $(1 + \epsilon)$.
Furthermore, we can assume without loss of generality that $C^\star$ is the
$(1+\epsilon)$-approximate solution claimed by \cref{dislabel},
Then by \cref{dislabel}, any distance $d(c, x)$ for $c \in C^\star$
and $x \in X$ has to lie in some interval
$\mathcal{I}_j = [\epsilon d_j, \epsilon^{-1} d_j]$,
and because of the rounding of distances,
the distances on $C^\star \times X$ have to take from a set
$\{r_1, \ldots, r_t\}$, where $t = \poly(\epsilon^{-1} \|X\|_0)$.

However, $C^\star$ is not known by the algorithm, and we have to ``guess'' $c$ and $c'$.
Specifically we enumerate over all points $x\in X$
which corresponds to the nearest point of $c$, and connection costs
$r\in \{r_1,...,r_t\}$ corresponding to $d(x, c)$,
where $c$ is some imaginary center in $C^\star$.
To implement this efficiently, we pre-process the distances on $V \times X$
using $O(\|X\|_0)$ runs of Dijkstra's algorithm in time $\tilde{O}(\|X\|_0|V|)$,
and then $r_i$'s are enumerated in $O(t)$ time.

Then to find $c'$, a naive approach is to add an $\epsilon r$-net of $B(x,r)$ into $D$.
The problem is that there may be too many points in the $\epsilon$-net,
so we need to use the structure of the graph to construct the net more carefully,
and we make use of \cref{lemma:planar_sep} which is restated as follows.
\begin{lemma}[Restatement of \cref{lemma:planar_sep}]
    \label{lemma:planar_sep_restate_centroid}
    For every edge-weighted planar graph $G = (V, E)$ and subset $S \subseteq V$,
    $V$ can be broken into parts $\Pi := \{ V_i \}_i$
    with $|\Pi| = \mathrm{poly}(|S|)$ and $\bigcup_{i}{V_i} = V$, such that
    for every $V_i \in \Pi$,
    \begin{enumerate}
        \item $|S \cap V_i| = O(1)$,
        \item there exists a collection of shortest paths $\calP_i$ in $G$ with $|\calP_i| = O(1)$ and
        removing the vertices of all paths in $\calP_i$ disconnects $V_i$ from $V\setminus V_i$ (points in $V_i$ are possibly removed).
    \end{enumerate}
    Furthermore, such $\Pi$ and the corresponding shortest paths $\calP_i$
    for $V_i \in \Pi$ can be computed in $\tilde{O}(|V|)$ time.
\end{lemma}
Apply \cref{lemma:planar_sep_restate_centroid} with (unweighted) $S = X$ to compute
parts $\Pi$ and the corresponding shortest paths $\calP_i := \{ P_j \}_j$
for each $V_i \in \Pi$, in $\tilde{O}(|V|)$ time.
Then, apart from enumerating $x_c$ and $r$, we further enumerate the set $V_i \in \Pi$.
For each $P^i_j \in \calP_i$, we let $Q^i_j := P^i_j\cap B(x_c,\epsilon^{-1}r+ r)$.
Observe that $P^i_j$ is a path,
so by triangle inequality $Q^i_j$ is contained in a segment of length
$O(\epsilon^{-1}r)$ of $P^i_j$.
We further find an $\epsilon r$-net\footnote{
For $\rho > 0$ and some subset $W \subseteq V$,
a $\rho$-net is a subset $Y \subseteq V$ such that
$\forall x, y \in Y$, $d(x, y) \geq \rho$ and $\forall x \in W$
there is $y \in Y$ with $d(x, y) < \rho$.
}
$R^i_j$ for $Q^i_j$ which is of size $O(\epsilon^{-2})$.
Finally, we let $R_i := \bigcup_j R^i_j$ denote the union of net points in
all the shortest paths in $\calP_i$, and $R'_i := R_i \cup (X\cap V_i)$
as the set with $X \cap V_i$ included in $R_i$.
By \cref{lemma:planar_sep_restate_centroid}, we know $|X \cap V_i| = O(1)$.
Write $R'_i = \{y_1, \ldots, y_m\}$.
We consider the set of possible distance tuples to $R'_i$,
i.e. for a point $x$, we consider the vector $(d(x,y_1), \ldots, d(x, y_m))$.

To restrict the number of possible distance tuples,
we need to carefully discretize the distances so that the distances only come from a small ground set.
\begin{itemize}
    \item For $y\in X\cap V_i$, because of \cref{dislabel}, we
    can discretize and assume $d(x,y)$ from $\{r_1,...,r_t\}$. 
    \item For $y \in R_i$, we note that we will only use $d(x, y)$
    such that $d(x, y) = O(r / \epsilon)$, so we only need to take
    $d(x,y)$ from $\{0,\epsilon r,2\epsilon r,...,(\epsilon^{-2}+5)\epsilon r\}$ (noting that here we use an $\epsilon r$ additive stepping).
\end{itemize}
Since $|R_i|=O(\epsilon^{-2})$ and $|X\cap V_i|=O(1)$,
there are $(\epsilon^{-2})^{O(\epsilon^{-2})}t^{O(1)}$ many possible tuples. 

For every tuple $(a_1, \ldots, a_m)$,
we find an arbitrary point $x'$ in $V_i \cap B(x_c, r)$ (if it exists) that realizes
the distance tuple to $R'_i$ when rounding to the closest discretized distance, i.e. $d'(x', y_i) = a_i$ for $1 \leq i \leq m$ where $d'$ is the discretized distance,
and add $x'$ into $D$.

In total, we have added
$(\epsilon^{-2})^{O(\epsilon^{-2})}t^{O(1)}\poly(\|X\|_0) =(\epsilon^{-1})^{O(\epsilon^{-2})}\poly(\|X\|_0) $ points into $D$, as desired.
This whole process of enumerating $V_i$, computing $\epsilon r$-nets
and finding point $x'$ for each tuple can be implemented in time $O( (\epsilon^{-1})^{O(\epsilon^{-2})}\poly(\|X\|_0) |V|)$.

\paragraph{Error Analysis}
We will prove $D$ is indeed an $\epsilon$-centroid set.
Consider the solution $C^\star = \{c_1, \ldots, c_k\}$ and the corresponding assignment $\pi$ guaranteed by \cref{dislabel}.
Suppose $C^\star$ clusters $X$ into $\{ C^\star_1, \ldots, C^\star_k \}$ by the arrangement $\pi$.
We will prove the following claim.
\begin{claim}
    \label{claim:approx_center}
    For every $1 \leq i \leq k$, there exists $c'_i \in D$ such that
    \begin{align}
        \forall y \in C^\star_i, \quad
        d(y, c'_i)\leq (1+O(\epsilon)) \cdot d(y, c_i), \label{eqn:approx_center}
    \end{align}
    where $C^\star_i \subseteq X$ is the cluster of $X$
    corresponding to $c_i \in C^\star$.
\end{claim}
Suppose the above claim is true,
then we define a $k$-subset $C' := \{c'_1, \ldots, c'_k\}$,
and it implies that $\cost(X, C') \leq (1 + O(\epsilon))  \cdot \cost(X, C^\star)$.
Hence, it remains to prove \cref{claim:approx_center},

\begin{proof}[Proof of \cref{claim:approx_center}]
Fix $1 \leq i \leq k$.
We start with defining $c'_i$.
Suppose $x_{c_i} \in X$
is the closest point to $c_i$ and let $r_i := d(x_{c_i}, c_i)$.
Let $V_j \in \Pi$ such that $c_i\in V_j$, and consider the moment
that our algorithm enumerates $x_{c_i}$, $r_i$ and $V_j$.
By construction, we have the following fact.
\begin{fact}
    \label{fact:centroid_construct}
There exists some point $c'\in D$
such that
    \begin{enumerate}
        \item $d(x_{c_i},c')\leq r_i$
        \item for every $y\in R_i$, if $d(c_i, y)\leq (\epsilon^{-2}+4)\epsilon r_i$, then $d(c', y)\in d(c_i, y) \pm \epsilon r_i$
        \item for every $y \in X \cap V_i$, $d(c',y)\in (1\pm\epsilon) \cdot d(c_i,y)$.
    \end{enumerate}
\end{fact}
We pick $c_i'$ as any of such $c'$ in \cref{fact:centroid_construct}.

Now we analyze the error. Fix $y \in C^\star_i$.
We note that the $R'_i$ that we pick only covers an $O(\epsilon^{-1} r_i)$ range,
so even though $c'_i$ approximate $c_i$ on the distance tuples,
it cannot directly imply the distance from $c'_i$ to
all other points in $C^\star_i$ is close to that from $c_i$,
and we need the following argument.

\begin{itemize}
    \item If $d(y, c_i)>\epsilon^{-1}r_i$, then $y$ is far away and
    $d(y, c'_i)$ cannot be handled by the distance tuples.
    However, we observe that in this case $d(c_i, c'_i)$ is small relative to $d(y, c_i)$.
    In particular, we have
    $d(c_i, c'_i) \leq d(c_i, x_{c_i}) + d(c'_i, x_{c_i}) \leq 2r_i$.
    Hence, it implies
    \begin{align*}
        d(y,c'_i)
        \leq d(y, c_i) + d(c_i, c'_i)
        \leq d(y, c_i) + 2 r_i
        \leq (1+2\epsilon) \cdot d(y, c_i).
    \end{align*}
    \item 
        Otherwise, $d(y, c_i)\leq \epsilon^{-1}r_i$,
        and we will use that $c'_i$ and $c_i$ are close with respect to the tuple distance,
        and use the separating shortest paths $\calP_j$
        (recalling that $V_j \in \Pi$ is the part that $c_i$ belongs to).
        \begin{itemize}
            \item If $y \in V_j$, then $y$ belongs to the set $R'_j$ and
            $d(c_i', y)$ belongs to one of the distance tuples
            (recalling that $y \in C^\star_i \subseteq X$).
            Hence, by the guarantee of the distance tuples,
            $c_i'$ satisfies $d(y, c_i') = d(y, c_i)$.
            \item Otherwise, $y \notin V_j$.
            Then the shortest path $y \rightsquigarrow c_i$ has to pass through
            at least one of the shortest paths in $\calP_j$.
            Now suppose $P^j_l \in \calP_j$ is the separating shortest path
            that shortest path $y \rightsquigarrow c_i$ passes through.
            Since $d(y, c_i) \leq \epsilon^{-1} r_i$,
            we have $y \in B(x_{c_i}, \epsilon^{-1} r_i + r_i)$.
            Since $P^j_l$ is a shortest path in $G$, $y \rightsquigarrow c_i$
            can only cross it once. 
        
            Hence,
            there is $y', y'' \in Q^j_l$ such that
            \begin{align*}
                d(y, c_i) = d(y, y') + d(y', y'') + d(y'', c_i).
            \end{align*}
            Since $R^j_l$ is an $\epsilon r_i$-net of $Q^j_l$, by triangle inequality, we know there exists $z',z''\in R^j_l$ such that
            \begin{align*}
            d(y, z') + d(z', z'') + d(z'', c_i)\leq d(y,c_i)+4\epsilon r_i.
            \end{align*}
            
Since $d(z'',c_i)\leq d(y,c_i)+4\epsilon r_i\leq (\epsilon^{-2}+4) \epsilon r_i$, by \cref{fact:centroid_construct} we know that
            $d(z'',c_i')\leq d(z'',c_i)+\epsilon r_i$. Finally, by triangle inequality, we have,
\begin{align*}
                d(y, c_i') \leq d(y, z') + d(z', z'') + d(z'', c_i')\leq d(y, c_i) + O(\epsilon r_i),
            \end{align*}
            
            Observe that by definition $d(y, c_i) \geq d(x_{x_i}, c_i) = r_i$,
            so we conclude that
            \begin{align*}
                d(y, c_i') \leq d(y, c_i) + O(\epsilon r_i)
                \leq (1 + O(\epsilon)) \cdot d(y, c_i).
            \end{align*}
        \end{itemize}
\end{itemize}
This completes the proof of \cref{claim:approx_center}.
\end{proof}
This completes the proof of \cref{thm:centroid_planar}
\end{proof}

\subsection{Improved PTAS's for Planar \kMedian}
\label{sec:ptas_planar}
Recently, \cite{cohen2019local} showed the local search algorithm
that swaps $\epsilon^{-O(1)}$ points in the center set
in each iteration yields a $1+\epsilon$ approximation for \kMedian
in planar and the more general excluded-minor graphs.
We use the centroid set and coreset to speedup this algorithm, and we
obtain the following PTAS.
\begin{corollary}
    \label{cor:planar_ptas}
    There is an algorithm that for every $0 < \epsilon < \frac{1}{2}$,
    integer $k \geq 1$ and every edge-weighted planar graph $G=(V, E)$,
    computes a $(1 + \epsilon)$-approximate solution for \kMedian
    on every weighted set $X \subseteq V$ with constant probability,
    running in time $\tilde{O}((\epsilon^{-1}k)^{\epsilon^{-O(1)}}\cdot |V|)$.
\end{corollary}
As noted by \cite{DBLP:conf/focs/HuangJLW18} and \cite{DBLP:journals/siamcomp/FriggstadRS19},
the potential centers that the local search algorithm should consider
can be reduced using an $\epsilon$-centroid set,
but to make the local search terminate properly,
we also need to evaluate the objective value accurately in each iteration
, which means we also need a coreset.
Hence, we start with constructing a coreset using \cref{cor:coreset_mf},
and then extend it to be a centroid set using \cref{thm:centroid_planar}.

\begin{proof}[Proof of \cref{cor:planar_ptas}]
Construct an $\epsilon$-coreset $S$ of size $\poly(\epsilon^{-1} k)$
using \cref{cor:coreset_mf},
and apply \cref{thm:centroid_planar} with $X = S$ to obtain
an $\epsilon$-centroid set $S'$ of size $(\epsilon^{-1} )^{O(\epsilon^{-2})}k^{O(1)}$.
Then the algorithm constructs a weighted set $D$ that consists of $S \cup S'$,
and the weights of points $x \in S$ is set to $w_S(x)$,
and those $x \in S' \setminus S$ has weight $0$.
It is immediate that $D$ is both an $\epsilon$-coreset and an $\epsilon$-centroid set, whose size is $\|D\|_0 = (\epsilon^{-1} )^{O(\epsilon^{-2})}k^{O(1)}$.
We pre-process the pairwise distance in $D \times D$ using Dijkstra's algorithm.
The overall running time for all these steps is
$\tilde{O}((\epsilon^{-1})^{\epsilon^{-O(1)}}k^{O(1)}\cdot |V|)$.

We next use $D$ to accelerate~\cite[Algorithm 1]{cohen2019local}.
The algorithm first defines an initial center set $C$ to be an arbitrary subset of $D$.
Then in each iteration, the algorithm enumerates $C' \in D^k$
that is formed by swapping at most $\epsilon^{-O(1)}$ points in $D$ from $C$.
Update $C := C'$ if some $C'$ has cost $\cost(D, C') \leq (1 - \frac{\epsilon}{|V|}) \cdot \cost(D, C)$,
and terminate otherwise.
The running time for each iteration is $(\epsilon^{-1} k)^{\epsilon^{-O(1)}}$.

By~\cite{cohen2019local}, the algorithm always finds a $(1 + \epsilon)$-solution when it terminates,
and the number of iterations is at most $\epsilon^{-1} |V|$ until termination. Therefore,
the total running time is bounded by
$\tilde{O}((\epsilon^{-1} k)^{\epsilon^{-O(1)}}\cdot |V|)$.
This completes the proof.
\end{proof}
 
\bibliographystyle{alphaurl}
\bibliography{main}

\appendix
\appendixpage
\section{Proof of \cref{lemma:planar_sep}}
\label{sec:proof_planar_sep}

\begin{lemma}[restatement of \cref{lemma:planar_sep}]
For every edge-weighted planar graph $G = (V, E)$ and subset $S \subseteq V$,
    $V$ can be broken into parts $\Pi := \{ V_i \}_i$
    with $|\Pi| = \mathrm{poly}(|S|)$ and $\bigcup_{i}{V_i} = V$, such that
    for every $V_i \in \Pi$,
    \begin{enumerate}
        \item $|S \cap V_i| = O(1)$,
        \item there exists a collection of shortest paths $\calP_i$ in $G$ with $|\calP_i| = O(1)$ and
        removing the vertices of all paths in $\calP_i$ disconnects $V_i$ from $V\setminus V_i$ (points in $V_i$ are possibly removed).
    \end{enumerate}
    Furthermore, such $\Pi$ and the corresponding shortest paths $\calP_i$
    for $V_i \in \Pi$ can be computed in $\tilde{O}(|V|)$ time.\end{lemma}

The proof of \cref{lemma:planar_sep} is based on the following property of general trees.
We note that the special case when $R = T$ was proved
in~\cite[Lemma 3.1]{DBLP:conf/soda/EisenstatKM14} and our proof is based on it. 
Nonetheless, we provide the proof for completeness.
\begin{lemma}\label{TreePartition}
    Let $T$ be a tree of degree at most $3$ and let $R$ be a subset of nodes in $T$.
    There is a partition of the nodes of $T$ with $\mathrm{poly}(|R|)$
    parts, such that each part is a subtree of $T$ that contains
    $O(1)$ nodes of $R$ and has at most
    $4$ boundary edges\footnote{Here a boundary edge is an edge that has exactly one endpoint in the subtree.} connecting to the rest of $T$.
    Such partition can be computed in time $\tilde{O}(|T|)$, where $|T|$ is the number of nodes in $T$.
\end{lemma}

\begin{proof}
    We give an algorithm to recursively partition $T$ in a top-down manner.
    The recursive algorithm takes a subtree $T'$ as input,
    and if $|T' \cap R| \geq 4$,
    it chooses an edge $e$ from $T'$ and run recursively on the two subtrees
    $T'_1$ and $T'_2$ that are formed by removing $e$ from $T'$.
    Otherwise, the algorithm simply declares the subtree $T'$
    a desired part and terminate, if $|T' \cap R| < 4$.
    Next, we describe how $e$ is picked provided that $|T' \cap R| \geq 4$.

    If $T'$ has at most $3$ boundary edges, we pick an edge $e \in T'$
    such that each of the two subtrees $T'_1$, $T'_2$ formed by removing $e$
    satisfies $\frac{1}{3} |T' \cap R| \leq |T'_j \cap R| \leq \frac{2}{3} |T' \cap R|$, for $j = 1, 2$.
    By a standard application of the balanced separator theorem
    (see e.g. Lemma 1.3.1 of~\cite{planarity}), such edge always exists
    and can be found in time $O(|T'|)$.

    Now, suppose $T'$ has exactly $4$ boundary edges.
    Then we choose an edge $e \in T'$, such that each of the two subtrees $T'_1$
    and $T'_2$ formed by removing $e$ has at most $3$ boundary edges.
    Such $e$ must exist because the maximum degree is at most $3$,
    and such $e$ may be found in time $O(|T'|)$ as well.
    To see this, suppose the four endpoints (in $T'$) of the four boundary edges are $a,b,c,d$.
    It is possible that they are not distinct, but they can
    have a multiplicity of at most $2$ because otherwise the
    degree bound $3$ is violated.
    If any point has a multiplicity $2$, say $a$ and $b$,
    then it has to be a leaf node in $T'$ (again, because of the degree constraint),
    and we can pick the unique tree edge in $T'$ connecting $a$
    as our $e$.
    Now we assume the four points are distinct,
    and consider the unique paths $P_1$, $P_2$ that
    connect $a,b$ and $c,d$ respectively.
    If $P_1$ and $P_2$ intersect, then the intersection must contain
    an edge as otherwise the intersections are at nodes only which means 
    each of them have degree at least $4$, a contradiction.
    Hence, we pick the intersecting edge as our $e$.
    Finally, if $P_1$ and $P_2$ are disjoint, we consider the unique path $P_3$
    that connects $a$ and $c$, and we pick edge $e := e'$ in $P_3$
    that is outside both $P_1$ and $P_2$ to separate $a$ and $b$ from $c$ and $d$.

    We note that there are no further cases regarding the number of boundary edges of $T'$,
    since in the case of $4$ boundaries edges,
    both $T'_1$ and $T'_2$ have at most $3$ boundary edges
    and it reduces to the first case.

    It remains to analyze the size of the partition.
    By the property of balanced separator, we know that such recursive partition has $O(\log |R|)$ depth.
    Hence the total number of subtrees is $2^{O(\log |R|)}=\mathrm{poly}(|R|)$. Finally, we note that in each level of depth, we scan the whole tree once, so the running time is upper bounded By $O(\log |R|)\cdot |T|=\tilde{O}(|T|)$.
\end{proof}

\begin{proof}[Proof of \cref{lemma:planar_sep}]
    We assume $G$ is triangulated, since otherwise
    we can triangulate $G$ and assign weight $+\infty$ to the new edges
    so that the shortest paths are the same as before.
    Let $T$ be a shortest path tree of $G$ from an arbitrary root vertex.
    Let $G^\star$ be the planar dual of $G$.
    Let $T^\star$ be the set of edges $e$ of $G^\star$ such that
    the corresponding edge of $e$ in $G$ is not in $T$.
    Indeed, $T$ and $T^\star$ are sometimes called
    \emph{interdigitating} trees,
    and it is well known that $T^\star$ is a spanning tree of $G^\star$
    (see e.g.~\cite{planarity}).

    Choose $R^\star$ to be the set of faces that contain at least one point from $S$.
    We apply \cref{TreePartition} on $R = R^\star$ and $T = T^\star$
    to obtain $\Pi^\star$, the collection of resulted subtrees of $T^\star$.
    Then $|\Pi^\star| = \poly(|S|)$, and each part $C^\star$ in $\Pi^\star$ is
    a subset of faces in $G$ such that only $O(1)$ of these faces
    contain some point in $S$ on their boundaries.
    For a part $C^\star$ in $\Pi^\star$,
    let $V(C^\star)$ be the set of vertices in $G$ that are contained
    in the faces in $C^\star$.
    Recall that $G$ is triangulated, so each face can only contain $O(1)$
    vertices from $S$ on its boundary.
    Therefore, for each part $C^\star$ in $\Pi^\star$, $|C^\star \cap S| = O(1)$.

    Still by \cref{TreePartition}, each part $C^\star$ in $\Pi^\star$
    corresponds to a subtree in $T^\star$,
    and it has at most $4$ boundary edges connecting to the rest of $T^\star$.
    By the well-known property of planar duality (see e.g.~\cite{planarity}),
    each $C^\star$ is bounded by the fundamental cycles in $T$ of the boundary edges.
    We observe that the vertices of a fundamental cycle lie on $2$ shortest paths
    in $G$ via the least common ancestor in $T$
    (recalling that $T$ is the shortest path tree).
    So by removing at most $8$ shortest paths in $G$,
    $V(C^\star)$ is disconnected from $V\setminus V(C^\star)$ for every $C^\star \in \Pi^\star$.

    Therefore, we can choose $\Pi := \{ V(C^\star) : C^\star \in \Pi^\star \}$. For the running time, we note that both the triangulation and the algorithm in \cref{TreePartition} run in $\tilde{O}(|V|)$ time.
    This completes the proof.
\end{proof} 
\fi

\end{document}